\documentclass[pra,amsmath,amssymb,amsfonts,superscriptaddress,twocolumn]{revtex4-2}
\usepackage{graphicx,color,mathtools,bm,braket,newtxtext,newtxmath}

\providecommand{\abs}[1]{\left\lvert#1\right\rvert}

\usepackage{tabularx}
\usepackage{lipsum} 
\usepackage{ulem}

\usepackage[hyphenbreaks]{breakurl}
\usepackage[colorlinks=true,linkcolor=blue,citecolor=blue,urlcolor =blue]{hyperref}

\usepackage{appendix}
\usepackage{amsfonts,amssymb}
\usepackage{physics}
\usepackage{tikz}

\usepackage[percent]{overpic}
\usepackage{subcaption}

\begin{document}

\author{Nicolas Fabre\footnote{nicolas.fabre@telecom-paris.fr}}
\affiliation{Télécom Paris-LTCI, Institut Polytechnique de Paris, 19 Place Marguerite Perey, 91120 Palaiseau, France}

\date{\today}
\begin{abstract}
The discrete cylinder phase space offers a natural setting for twisted photons, Josephson junctions, and certain bosonic codes. In that context, we build a number-of-zeros hierarchy for the stellar distribution of states on the discrete cylinder phase space, tied to usefulness for quantum computation. The orbital angular momentum (OAM) and phase eigenstate, having no zeros, sits at the bottom of the hierarchy and share mathematical properties with the coherent state on the plane phase space, both sharing zero rank and a positive Wigner distribution. This reveals a structural peculiarity of the cylinder: its coherent state---the wrapped coherent state---is instead the most quantum, carrying infinitely many zeros, so "coherent state" need not signal classicality across different phase-space geometries. We identify the operations that preserve or change the zero count, giving explicit access to every rank of the hierarchy. We then ask whether this hierarchy also governs localization. Using the $L^r$ norm, the inverse participation ratio, and the Wehrl entropy as complementary measures of spread in the Husimi density, we find that the two hierarchies need not track one another: the wrapped coherent state can be more localized than the OAM eigenstate even though it carries infinitely many zeros. 
\end{abstract}
\vskip2pc 

\title{On the zeros of the stellar representation in the discrete cylinder}
\maketitle

\section{Introduction}

A quantum state can be described either by a density matrix or by a quasiprobability distribution in phase space, a framework that bridges quantum and classical statistical mechanics~\cite{Cahill:1969,Glauber:1963}. The best-known such distribution, the Wigner function, was introduced by Wigner in his seminal 1932 paper~\cite{Wigner:1932}. An especially appealing formulation for finite-dimensional systems is instead provided by the Majorana stellar representation, which maps pure spin states onto a constellation of points on the Bloch sphere~\cite{Majorana:1932ul,bruno_quantum_2012,Bjork:2015ux,Liu:2017,Chryssomalakos:2018}. This geometric picture not only gives an intuitive visualization of quantum states but also introduces the concept of trajectories -- reminiscent of classical motion -- into the quantum framework~\cite{Bacry:2004aa}. Recent work has extended the Majorana stellar representation beyond the Bloch sphere to infinite-dimensional systems, such as the complex plane~\cite{Chabaud:2020th} and the discrete cylinder~\cite{Gazeau:2022aa,PhysRevResearch.5.L032006}. The latter is particularly significant, as it is naturally the phase space of the canonical pair formed by the angle and the orbital angular momentum~\cite{Kastrup:2016,Rigas:2017}, and it is not merely of theoretical interest. Twisted photons~\cite{molina-terriza_twisted_2007,Yao:11,forbes_quantum_2019}, atomic vortex states in a sodium Bose--Einstein condensate~\cite{PhysRevLett.97.170406,PhysRevA.74.053809}, Josephson junctions~\cite{dirienzo_coupled_1983,fan_cooper-pair_2006}, and exciton-polariton condensates~\cite{ma_realization_2020,sedov_circular_2021} are all effectively described by such a phase-space distribution. Accordingly, the Wigner function has itself been extended both to the discrete cylinder $\mathbb{S}^{1}\times\mathbb{Z}$~\cite{mukunda_wigner_1979,Rigas:2010um,Rigas:2011,Aremua:2012,Fresneda:2018,fabre_wigner_2020,Kowalski:2021,Gazeau:2022aa}, measured experimentally in \cite{PhysRevLett.116.130402} and to its continuous, Euclidean counterpart $\mathbb{S}^{1}\times\mathbb{R}$~\cite{nieto_wigner_1998,Kastrup:2016,PhysRevA.95.052111}.

In continuous-variable quantum systems, non-Gaussianity plays a critical role in surpassing classical limits in quantum computing~\cite{Walschaers:2017,Takagi:2018,Albarelli:2018,Chabaud:2021,chabaud_resources_2023,descamps_superselection_2024}. A promising approach to quantifying non-Gaussianity is through the stellar representation, specifically, by analyzing the zeros in the Husimi $Q$-function~\cite{Chabaud:2020th}. In this framework, Gaussian states, which lack any zeros, form the base of a hierarchy, while states with increasing number of zeros is an indicator of non-classicality~\cite{walschaers_non-gaussian_2021,5fl9-89j4}. For example, the photon-addition process applied to the vacuum produces a Fock state, whose Husimi function acquires a single zero, marking the first level of non-Gaussian behavior~\cite{Chabaud:2020th,walschaers_non-gaussian_2021}. This connection between zeros of the Husimi function and non-Gaussianity is rooted in the equivalence, for pure states, between Husimi zeros and Wigner function negativity~\cite{Lutkenhaus:1995}. Moreover, pure Gaussian states are characterized by a strictly positive Wigner distribution, a result known as Hudson's theorem~\cite{Hudson:1974,Soto:1983}, together with the trace identity $\operatorname{Tr}(\hat\rho\,\hat\rho')=\int_{-\pi}^{\pi}\! d\phi\sum_{\ell\in\mathbb Z}W_{\hat\rho}(\ell,\phi)\,W_{\hat\rho'}(\ell,\phi)$ which enables their efficient classical simulation~\cite{PhysRevLett.109.230503}. In contrast, highly non-classical states, such as ideal Schr\"odinger cat states and Gottesman-Kitaev-Preskill (GKP) states, exhibit an infinite number of zeros, underscoring their potential as indispensable resources for fault-tolerant universal quantum computation~\cite{gottesman_encoding_2001,PhysRevLett.109.230503,Baragiola:2019,Yamasaki:2020,Bourassa:2021,Noh:2022,Royer:2020,Mirrahimi:2014,Rosenblum:2018,Schlegel:2022,Albarelli:2018}.

In the discrete cylinder many definitions of coherent states exist, each possessing different properties of the canonical $H(3)$ coherent states~\cite{Gazeau:2022aa,Kastrup:2016}. Since the stellar distribution is defined as the overlap between a wavefunction and a coherent state, its properties depend critically on which coherent state is chosen. In~\cite{PhysRevResearch.5.L032006}, the stellar distribution into the discrete cylinder of the wrapped coherent state, an eigenstate of an annihilation operator of a ladder operator into the $e(2)$ algebra \cite{Kowalski:1996mz}.

In this paper we establish a number-of-zeros hierarchy of the stellar distribution on the discrete cylinder phase space. This hierarchy reflects their usefulness for quantum computation. The most classical states are the orbital angular momentum (OAM) eigenstates, but also phase eigenstate, which sit at the bottom of the hierarchy, with no zero at all. We show that operators built from the ladder operator into the $e(2)$ algebra, one cans place a zero at any prescribed point of the cylinder, giving explicit access to every intermediate rank. Conversely, the coherent states of the non-compact Euclidean group realized on the discrete cylinder, for instance the wrapped coherent state is the most quantum: their stellar representations carry infinitely many zeroes, which for the wrapped coherent state we show sit on a single line, diametrically opposite its peak. This is corroborated by the sign of the Wigner distribution, positive for the OAM and phase eigenstate and negative for the wrapped coherent state \cite{Rigas:2010um}. It is worth emphasizing that a state can combine no zeroes with a positive Wigner function. Unlike the continuous-variable case, the discrete cylinder lacks a smoothing relation between Wigner and Husimi distributions, so these properties are logically independent, though they coincide here. Comparing the cylinder’s and plane’s zero-counting hierarchies suggests a partial OAM eigenstate-coherent state correspondence, extending to infinite-rank states: sums of OAM eigenstates yield wrapped coherent states, just as coherent sums of squeezed states yield GKP states, that is a correspondence we make precise by matching generating relations term-by-term. We then identify the set of operations that conserves and modifies the number of zeroes, similarly to the Gaussian operations for single mode continuous variables states.

It is natural to ask whether this zero-counting hierarchy also governs the localization of a state in phase space. It largely does not. To address this, we introduce several complementary quantifiers, the $L^{r}$ norm, the inverse participation ratio (IPR), and the Wehrl entropy, built from the multipole expansion in the discrete cylinder \cite{PhysRevResearch.5.L032006},  and use them to compare the OAM eigenstate, the wrapped coherent state, and the phase eigenstate, together with their associated superpositions. None of these metrics probes the location of the zeroes directly; rather, they measure the spread of the Husimi density. Phase space localisation and spreadness of zeros do not align: a wrapped coherent state, despite its infinite zeroes, has strongly damped zeroes in its Husimi distribution and can be more localized than an OAM eigenstate or finite-rank superposition. While the stellar representation of phase eigenstate occupies an infinite volume in phase space, and do not have any zeros.

The paper is organized as follows. In Sec.~\ref{establishingOAM} we recall the stellar distribution on the discrete cylinder, uncover further properties, and establish the hierarchy of OAM states by their number of zeros. We identify the set of operations that conserves the number of zeroes. In Sec.~\ref{OAMsecdis} we go further and introduce quantifiers of localizability on the discrete cylinder, the $L^{r}$ norm, the inverse participation ratio, and the Wehrl entropy, and use them to characterize the OAM eigenstate, the wrapped coherent state, and the phase eigenstate, and their superpositions.

\section{Establishment of the number of zeroes hierarchy in the cylinder}\label{establishingOAM}

\subsection{Algebra and coherent state in the discrete cylinder}
Following \cite{PhysRevResearch.5.L032006}, we recall the main properties of OAM states, and how to introduce coherent states for the discrete cylinder. The commutation between the OAM operator $\hat{L}$ and its conjugate angle $\hat{\phi}$ is expressed through the commutator~\cite{Lerner:1968aa,Hradil:2006aa}
\begin{equation}
[ \hat{E},  \hat{L}] = \hat{E} \, ,
\end{equation}
with $\hat{E} = e^{i \hat{\phi}}$, which characterizes the Euclidean Lie algebra $\mathfrak{e}_{2}$~\cite{Humphreys:1972aa}. With the cosine operator $\hat{H}=\frac{1}{2}k(\hat{E}+\hat{E}^{\dagger})$ and the sine operator $\hat{P}=-i\frac{1}{2}k(\hat{E}-\hat{E}^{\dagger})$, the Casimir operator defined as $\hat{K}^{2}=\hat{H}^{2}+\hat{P}^{2}=k^{2}\mathbb{I}$ has eigenvalue $k^{2}$ labeling the irreducible representations of $E(2)$~\cite{Nieto:1998aa}. Two conjugate bases are particularly relevant here:
\begin{equation}
\ket{\phi}=\frac{1}{\sqrt{2\pi}} \sum_{\ell \in\mathbb{Z}} e^{-i\ell \phi} \ket{\ell}, \quad
\ket{\ell} = \frac{1}{\sqrt{2\pi}} \int_{0}^{2\pi} \! \! d\phi \; e^{i\ell \phi} \ket{\phi} \, ,
\end{equation}
normalized according to $\bra{\ell}\ket{\ell'} = \delta_{\ell \ell^{\prime}}$ and $\bra{\phi}\ket{\phi'}=\delta_{2\pi}(\phi-\phi^\prime)=\sum_{n \in\mathbb{Z}} \delta(\phi-\phi^{\prime}-2 n \pi)$. One readily checks that $\ket{\ell}$ and $\ket{\phi}$ are eigenvectors of the corresponding operators, $\hat{L} \ket{\ell} = \ell \ket{\ell}$ and $\hat{E} \ket{\phi}= e^{i\phi}\ket{\phi}$. Writing the components of a pure state $\ket{\psi}$ in each basis as $\psi_{\ell} = \langle \ell | \psi \rangle$ and $\psi(\phi) = \langle \phi | \psi \rangle$, the two are related by a Fourier transform, reflecting the complementarity of $\hat L$ and $\hat\phi$. A displacement operator on the circle can then be defined, up to ordering conventions and phase factors, as
\begin{equation}
   \hat{D}(m,\phi)=e^{-i m\phi/2}\,\hat{E}^m\,\hat{V}(\phi), \qquad \hat V(\phi)=e^{-i\phi\hat L}, 
\end{equation}
so that $\hat D(m,\phi)$ displaces a state simultaneously in angular momentum (by an integer $m$) and in angle (by $\phi$). These operators obey a Weyl-type commutation relation,
\begin{equation}
\hat{D}(m_1,\phi_1)\hat{D}(m_2,\phi_2)=e^{-i(m_1\phi_2-m_2\phi_1)/2}\,\hat{D}(m_1+m_2,\phi_1+\phi_2),
\end{equation}
the group law of the Heisenberg--Weyl group adapted to the topology of the discrete cylinder.\\

 The covariant integral quantization technique (see Appendix \ref{app:covariant}) allows one to find different sets of coherent states on the discrete cylinder. A particular example of a coherent state for such a phase space is the wrapped coherent state, which can be written as~\cite{Kowalski:1996mz,Gazeau:2022aa,PhysRevResearch.5.L032006}
\begin{equation}
\label{eq:cscyl}
 |\ell,\phi\rangle=\frac{e^{-\ell^2/2}}{\sqrt{\varpi}} \sum_{n\in \mathbb{Z}}e^{-n^{2}/2} e^{n(\ell - i \phi)}|n \rangle \, ,
\end{equation}
with normalization $\varpi= \sum_{n \in\mathbb{Z}} e^{-n^2}\approx 1.776372$. In the $\phi$-representation,
\begin{equation}
    \ket{\ell,\phi}=e^{-\ell^{2}/2} \int_{0}^{2\pi} d\phi' \, \vartheta_{3}\!\Big({-}\frac{i\ell}{2}+\frac{1}{2}(\phi'-\phi),\,e^{-1/2}\Big) \ket{\phi'} \, ,
\end{equation}
where $\vartheta_{3}(z,q) = \sum_{n\in\mathbb{Z}} q^{n^2} e^{2inz}$ is the third Jacobi function. Unlike its $H(3)$ counterpart, the wrapped coherent state is $2\pi$-periodic. These states resolve the identity,
\begin{equation}
\label{eq:resun}
\frac{1}{2\pi}\sum_{\ell\in \mathbb{Z}}\int_0^{2\pi} d\phi \; |\ell,\phi\rangle\langle \ell,\phi| =\openone \,,
\end{equation}

A complex coordinate on the cylinder is obtained directly from $(\ell,\phi)$ through the linear identification:
\begin{equation}\label{conformalmapping}
z \equiv \ell - i\phi \in\mathbb C,
\end{equation}
so that a full turn $\phi\to\phi+2\pi$ corresponds to the translation $z\to z+2\pi i$: the discrete cylinder is thereby identified with an infinite strip of the complex plane, periodically repeated along the imaginary direction. The multiplicative eigenvalue of the ladder operator $\hat X$ introduced below is $e^{z}=e^{\ell-i\phi}$.

The operators acting on these states are the ladder operators $\hat{X}=e^{\hat{L}-i\hat{\phi}}$ and $\hat{X}^{\dagger}$~\cite{Kowalski:1996mz},
\begin{align}
\hat{X}\ket{l}=e^{-l-1/2}\ket{l+1}\\
\hat{X}^{\dagger}\ket{l}=e^{-l+1/2}\ket{l-1} \, ,
\end{align}
where $[\hat{X},\hat{X}^{\dagger}]=2\text{sinh}(\mathbb{I})e^{-2\hat{L}}$. The wrapped coherent state $\ket{z}$ is the eigenstate of $\hat{X}$~\cite{Kowalski:1996mz,PhysRevResearch.5.L032006}, $\hat{X}\ket{z}=e^{z}\ket{z}$; together with the relations above, this identifies $\hat X$ as an annihilation-type operator. This is the only property that the wrapped coherent state shares with the coherent state $\hat{\alpha}$ on the plane: $\hat{a}\ket{\alpha}=\alpha \ket{\alpha}$.

We now introduce the conformal transformation $\xi=e^{z}=e^{l-i\phi}$, where $\xi$ corresponds to the coordinate into a punctured plane. The negative infinite in the cylinder corresponds to the origin of the punctured plane, while the positive infinite of the cylinder correspond to a circle of infinite radius. We provide in Appendix \ref{sec:conformal} more details about such a conformal transformation, and the issues of the singularity at the origin of the plane. Such a transformation plays a role for the choice of the parametrization of the stellar representation.

\subsection{Stellar representation in the discrete cylinder}
By contrast with \cite{PhysRevResearch.5.L032006}, we build differently the stellar representation to highlight that it can be built from the ladders operators into the $e(2)$ algebra.
Expanding the wrapped coherent state in the OAM basis $\{\ket l\}$,
\begin{equation}
\ket{z}=\sum_{l\in\mathbb{Z}} \braket{l}{z}\ket{l},
\end{equation}
with overlap $\braket{l}{z}=e^{-l^2/2} e^{-lz} \braket{0}{z}=e^{-l^2/2}  \bra{0}\hat{X}^{-l}\ket{z}$, where $\bra{0}\ket{z}=e^{-\abs{z}^{2}/2}/\sqrt{\varpi}$ gives
\begin{equation}
\ket{l}=e^{-l^{2}/2} (\hat{X}^{\dagger})^{-l} \ket{0} \, .
\end{equation}
Equivalently it can be written as $\ket{l}=e^{l^{2}/2} \hat{X}^{l}\ket{0}$
This writing is in strong mathematical analogy with Fock states: $\ket{n}=(\hat{a}^{\dagger})^{n}/\sqrt{n!} \ket{0}$. Any wavefunction decomposes in the OAM basis as
\begin{equation}\label{statedecomposition}
\ket{\psi}=\sum_{n\in\mathbb{Z}}\psi_{n} \ket{n}=\Big(\sum_{n\in\mathbb{Z}} \psi_{n} e^{-n^{2}/2} (\hat{X}^{\dagger})^{-n}\Big)\ket{0} \, ,
\end{equation}
such that $\sum_{n\in\mathbb{Z}}\abs{\psi_{n}}^{2}=1$ and the stellar function is defined directly as the overlap with the coherent state:
\begin{equation}\label{eq:stellar}
F_{\psi}(z)= \frac{e^{-\abs{z}^{2}/2}}{\sqrt{\varpi}}\sum_{n\in\mathbb{Z}} e^{-n^{2}/2} e^{nz} \psi_{n} \, .
\end{equation}
with $\varpi=\sum_{n\in\mathbb{Z}} e^{-n^2}\approx1.776372$ and where we will note $\frac{e^{-\abs{z}^{2}/2}}{\sqrt{\varpi}} G_{\psi}(z)\equiv\braket{z}{\psi}$. The wavefunction is generated from the vacuum as
\begin{equation}\label{wavefunction}
\ket{\psi}=\frac{e^{-\abs{z}^{2}/2}}{\sqrt{\varpi}} G_{\psi}\big({-}\ln\hat{X}^{\dagger}\big)\ket{0} \, .
\end{equation}
This mirrors the stellar formalism for continuous-variable states~\cite{Chabaud:2020th}, where a general state is built by repeated photon addition on the vacuum. Here, however, the OAM basis has no privileged origin analogous to the Fock vacuum, and the construction can equally start from any OAM state $\ket{j}$ -- a fact confirmed below by the zero-counting of Fock and OAM states. We further note that if $\ket{\phi}$ and $\ket{\psi}$ are normalized pure states with $G_{\phi}=G_{\psi}$, then $\phi_{n}=\psi_{n}$ for all $n$, since $G_\phi$ and $G_\psi$ are entire functions (see below) whose Taylor coefficients coincide.

Note that, the stellar representation into the $\xi$ coordinate on the punctured plane is 
\begin{equation}
    H_{\psi}(\xi)= \sum_{n\in\mathbb{Z}} e^{-n^{2}/2} \xi^{n} \psi_{n}.
\end{equation}
While $G_\psi(z)$ is an entire function of $z$ (see Appendix~\ref{app:sing}), the punctured-plane representation $H_\psi(\xi)$ is in general not entire because of singularities at $\xi=0$ and $|\xi|\to\infty$ arising from the infinite extent of the cylinder. For finite-rank states the sum truncates to a finite number of terms, so $H_\psi(\xi)$ becomes a Laurent polynomial and is entire up to a possible pole at the origin (absent when the state is such that $\psi_{0}=0$). The coefficients are recovered from a Fourier-type integral over one period,
\begin{equation}\label{Laurent}
\psi_{n}e^{-n^{2}/2}=\frac{1}{2i\pi} \oint_{\gamma} \frac{H_{\psi}(\xi)}{\xi^{n+1}}\, d\xi= \frac{1}{2i\pi} \oint_{\gamma} \frac{G_{\psi}(z)}{e^{zn}}\, dz  \, ,
\end{equation}
with $\gamma$ any contour enclosing a period strip of $2\pi i$. For infinite-rank states the contour must avoid the essential singularities at the punctured plane ends; this is achieved by excising small disks $|\xi|<\epsilon$ and $|\xi|>1/\epsilon$ and taking $\epsilon\to0$ after renormalisation of the relevant observables (see Appendix~\ref{app:sing}). For finite-rank states no such regularisation is needed and the integration can be performed directly.

\medskip
The Cauchy--Schwarz inequality, together with $\sum_{n\in\mathbb{Z}}|\psi_n|^2=1$, bounds the stellar function as
\begin{equation}
    \abs{G_\psi(z)}^{2} \leq \sum_{n\in\mathbb{Z}} e^{-n^{2}} e^{2n\,\mathrm{Re}(z)}
    \le \varpi\, e^{(\mathrm{Re}\,z)^{2}} \, ,
\end{equation}
the bound being saturated whenever $\mathrm{Re}(z)\in\mathbb Z$. Together with the absolute convergence of the defining series for every finite $z$ (Appendix~\ref{app:sing}), this shows that $G_\psi$ is an entire function of $z$, of order 2: its growth is Gaussian in $\mathrm{Re}(z)$, tracking the two ends $\ell\to\pm\infty$ of the cylinder. By the Hadamard factorization theorem for order-2 entire functions \cite{leboeuf_chaos-revealing_1990} (see Appendix~\ref{sec:bargmann}),
\begin{equation}\label{polynomial}
G_{\psi}(z)=e^{\,a+bz+cz^{2}}\prod_{i}E_2(z/z_{i})^{m_{i}}, \ E_2(u)=(1-u)\,e^{u+u^2/2} \, ,
\end{equation}
where the product runs over all zeros $z_i$ (finite or infinite in number), $m_i$ are their multiplicities, and $a,b,c$ are constants fixed by the state. For the sake of completeness, for more properties on the stellar representation into the $\xi$ coordinates in Appendix \ref{app:counting}.

The Husimi distribution on the discrete cylinder,
\begin{equation}
Q_{\psi}(z)=\abs{F_\psi(z)}^{2}, \qquad \int dz\, Q_{\ket\psi}(z)=1 \, ,
\end{equation}
shares the zeroes of $F_\psi$ and $G_\psi$, each with double multiplicity.

\subsection{Transformations that conserve the number of zeroes - states equivalence}

In this section we identify the operations that conserve the number of zeroes of the stellar function on the discrete cylinder.

Let us first recall the corresponding result for continuous-variable states. The vacuum, and more generally any pure Gaussian state or coherent state, has null stellar rank and sits at the bottom of the hierarchy; Gaussian operations do not change the number of zeroes, and Gaussian states can be simulated efficiently, in polynomial time, on a classical computer. The Fock states $\ket{n}$ (other than the vacuum) are non-Gaussian and possess $n$ zeroes.

We now identify the states that possess no zero on the cylinder. Since $G_\psi$ is entire and $2\pi i$-periodic in $z=\ell-i\phi$, the zero count is unambiguous: it is simply the number of zeroes of $G_\psi(z)$ in one period strip. For the OAM eigenstate $\ket{\ell_0}$, \textit{i.e}.\ $\psi_n=\delta_{n,\ell_0}$,
\begin{equation}
G_{\ell_0}(z)=e^{-\ell_0^{2}/2}\,e^{\ell_0 z},
\end{equation}
a nowhere-vanishing entire function of $z$. We point out that $H_{\ell_0}(\xi)=e^{-l_{0}^{2}/2} \xi^{l_{0}}$ which is however vanishing at $\xi=0$ and corresponds to a singularity of the conformal transformation (as it is mapped at the negative infinite of the cylinder).  Every OAM eigenstate therefore has exactly zero zeroes on the cylinder, independently of $\ell_0$, recovering the observation of~\cite{PhysRevResearch.5.L032006} directly, with no need to relate a punctured conformal plane back to the cylinder. We note that the stellar representation of the phase eigenstate does not have any zero \cite{PhysRevResearch.5.L032006}, while is delocalized in phase space.

Displacement operators do not change this count either. From the ladder relations for instance $\hat{X}\ket{j}=e^{-j-1/2}\ket{j+1}$ and $\hat{X}^{\dagger}\ket{j}=e^{-j+1/2}\ket{j-1}$ one finds, by direct computation (see \cite{Kowalski:1996mz} and Appendix \ref{sec:cylindrical} for supporting calculations),
\begin{equation}\label{eq:Xdagger-action}
G_{\hat X^{\dagger}\psi}(z)=e^{-z}\,G_\psi(z), \qquad
G_{\hat X\psi}(z)=e^{\,z-1}\,G_\psi(z-2) \, ,
\end{equation}
and, for the jump operator $\hat{E}=e^{i\hat\phi}$ (with $\hat E\ket n=\ket{n-1}$) and its inverse $\hat E^{-1}$,
\begin{equation}\label{eq:E-action}
G_{\hat E\psi}(z)=e^{-z-1/2}\,G_\psi(z+1), \quad
G_{\hat E^{-1}\psi}(z)=e^{\,z-1/2}\,G_\psi(z-1) \, .
\end{equation}
In every case the new stellar function is the old one, translated and multiplied by a nowhere-vanishing exponential prefactor: displacement operators can only relocate existing zeroes, never create or destroy one. More generally, any operator built from $\hat E^{m}$ conjugated by a function of $\hat L$ alone,
\begin{equation}\label{operatorconserve}
e^{i\nu \hat{L}^{k}/2} \, \hat{E}^{m} \, e^{-i\nu \hat{L}^{k}/2} \, , \qquad \nu\in\mathbb C, \; k\in\mathbb N,
\end{equation}
conserves the number of zeroes of the stellar distribution (Appendix~\ref{sec:conservation}); in particular the twist operator $\hat{T}(\nu)=e^{i\nu\hat{L}^{2}/2}$~\cite{potocek_exponential_2015} does so, operator which is useful for the indirect tomography of twisted photons~\cite{Rigas:2011ut}. These operators are, in this sense, the counterpart on the cylinder of the Gaussian operations for continuous variables system: acting with Eq.~\eqref{operatorconserve} on $\ket{\ell_0}$ produces another zero-free state, and more generally maps each rank of the hierarchy into itself.

\subsection{Transformations that increase the number of zeroes}

Equation~\eqref{eq:Xdagger-action} shows that $\hat{X}^{\dagger}$ acts on the stellar function by plain multiplication, $G_\psi(z)\mapsto e^{-z}G_\psi(z)$: since $e^{-z}$ never vanishes, $\hat X^\dagger$ alone can only move a state along the zero-free OAM sequence, exactly as observed above. Genuine zeroes are created by combining $\hat{X}^{\dagger}$ with the identity. For a target zero at $z_0\in\mathbb C$ define
\begin{equation}\label{cylinderjump}
\hat{O}(z_{0}) \equiv \hat{X}^{\dagger}-e^{-z_{0}}\,\mathbb{I} \, ,
\end{equation}
so that, starting from $\ket{0}$ (for which $G_{\ket 0}\equiv1$),
\begin{equation}
G_{\hat O(z_0)\ket 0}(z)=e^{-z}-e^{-z_{0}} \, ,
\end{equation}
an entire function with exactly one zero per period, at $z=z_0 \pmod {2\pi i}$. Reading off the coefficients of Eq.~(\ref{Laurent}) gives the corresponding (unnormalized) state,
\begin{equation}
\ket{\psi}= e^{1/2}\ket{-1}-e^{-z_{0}}\ket{0} \, .
\end{equation}
Because $\hat X^\dagger$ acts by multiplication, this construction composes trivially: for $N$ prescribed zeroes at $z_1,\dots,z_N$, we have $\hat{O}(z_{1},\dots,z_{N})
\equiv \prod_{i=1}^{N}\hat{O}(z_i)$ so that
\begin{equation}
G_{\hat O(z_1,\dots,z_N)\ket 0}(z)
=
\prod_{i=1}^{N}\bigl(e^{-z}-e^{-z_i}\bigr).
\end{equation}
with no cross-term bookkeeping required. Expanding the product in elementary symmetric functions of $\{e^{-z_i}\}$ and matching powers of $e^{-z}$ gives the state directly,
\begin{equation}
\ket{\psi}=\sum_{k=0}^{N}(-1)^{N-k}\,e^{k^{2}/2}\,e_{N-k}\!\big(e^{-z_{1}},\dots,e^{-z_{N}}\big)\,\ket{-k} \, ,
\end{equation}
$e_j$ denoting the $j$-th elementary symmetric polynomial ($e_0\equiv1$); for $N=1,2$ this reproduces the states above and their two-zero generalization, now without solving for auxiliary parameters.

For an $n$-fold zero at a single point $z_0$, $\hat O(z_0)^n\ket 0$ gives $G(z)=(e^{-z}-e^{-z_0})^n$ and
\begin{equation}\label{statejump}
\ket{z_{0}^{n}}=\big(\hat O(z_0)\big)^{n}\ket{0}=\sum_{k=0}^{n} \binom{n}{k} (-1)^{n-k}\, e^{k^{2}/2}\, e^{-(n-k)z_{0}} \,\ket{-k} \, ,
\end{equation}
the Gaussian weight $e^{k^2/2}$ being fixed by the normalization $\ket{-k}=e^{-k^{2}/2}(\hat X^{\dagger})^{k}\ket 0$ established above and required for Eq.~\eqref{statejump}.\\

 We discussed in more details in the Appendix \ref{app:modzero} other set of mathematical transformations that modify the number of zeroes.

  \begin{figure*}
 \begin{center}
\includegraphics[width=0.8\textwidth]{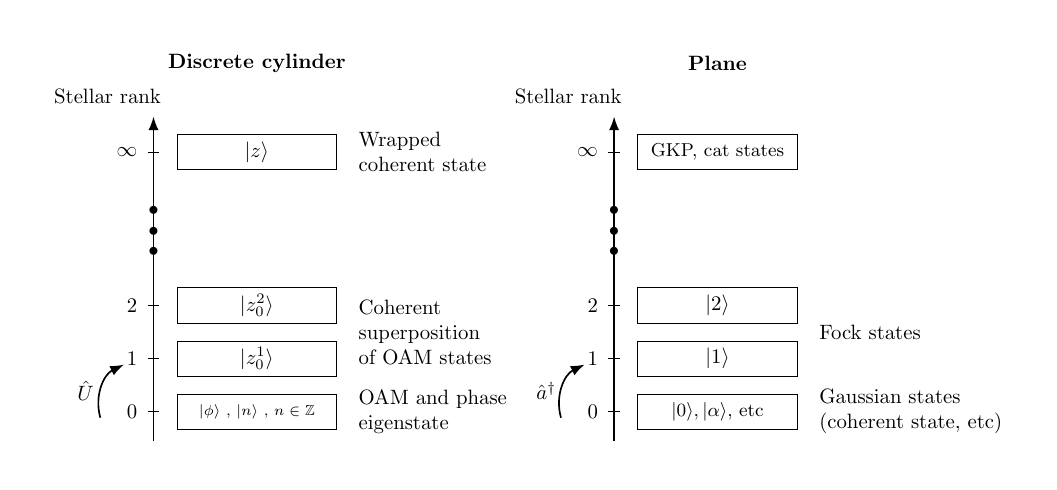}
\caption{\label{hierarchy}Stellar rank hierarchy for the discrete cylinder and plane. The stellar distribution of the OAM and phase eigenstates do not have any zeroes. The application of the jump operator $\hat{U}$ allow creating a zero not located at the origin of the plane; so that not at the negative infinite of the cylinder. For the notation $\ket{z_{0}^{n}}$ see Eq.~(\ref{statejump}), that corresponds to superposition of OAM eigenstates. Many possible states have the same rank. In the plane, the application of the annihilation or creation operator create or destroy one zero at the origin. However, for the discrete cylinder-conformal plane, ladders operators create or destroy one zero at the origin of the conformal plane, which correspond to a zero in the negative infinite in the discrete cylinder. Therefore, the application of the superposition of a ladder operator and the identity allow creating one zero located in the conformal plane except the origin, and therefore in the discrete cylinder.  }
\end{center}
\end{figure*}

\subsection{Full number of zeroes hierarchy in the discrete cylinder}

 We now assemble the full hierarchy of stellar rank on the discrete cylinder from the results above (Fig.~\ref{hierarchy}). Its bottom is occupied by all the OAM eigenstates, whose stellar functions  having no zero and a sharply peaked multipole expansion \cite{PhysRevResearch.5.L032006}, confirms that the OAM eigenstate is the most classical state on the cylinder: OAM eigenstates occupy the bottom of the stellar-rank hierarchy.  The OAM eigenstate $\ket{\ell_{0}}$ has a positive Wigner distribution~\cite{Rigas:2011ut}, and, together with the property $\text{Tr}(\hat{\rho}\hat{\rho}')=\int_{-\pi}^{\pi} d\phi \sum_{\ell\in\mathbb{Z}} W_{\hat{\rho}}(\ell,\phi) W_{\hat{\rho}'}(\ell,\phi)$, indicates that a circuit built solely from such states can be simulated efficiently on a classical computer~\cite{PhysRevLett.109.230503}. The phase eigenstate has also a positive Wigner distribution \cite{Rigas:2010um}, and the stellar distribution has no zeros (see \cite{PhysRevResearch.5.L032006} and also next section). Therefore, phase eigenstate sits also at the bottom of this hierarchy.  \\

Now, applying the operator of Eq.~\eqref{cylinderjump} to an OAM eigenstate adds exactly one zero, at the prescribed finite point $z_0$, moving the state up one rank; iterating, or combining several factors as above, populates the intermediate ranks of the hierarchy.\\

At the top sits the wrapped coherent state, which possesses infinitely many zeroes on the cylinder~\cite{PhysRevResearch.5.L032006}. In the present coordinates it is generated from the fiducial state $\ket\Omega=\sum_{l\in\mathbb Z}e^{-l^2/2}\ket l$ -- itself the wrapped coherent state at $z=0$, up to normalization -- by
\begin{equation}\label{wrappednumber}
\ket{z}\propto e^{z\hat{L}}\ket{\Omega} \, ,
\end{equation}
consistent with Eq.~\eqref{eq:cscyl}: since $\hat L\ket l = l\ket l$, this reproduces $e^{-l^2/2}e^{lz}$ term by term, and correspondingly $G_{z_0}(z)=G_\Omega(z+z_0)$ for the state centered at $z_0$. As shown in Appendix~\ref{app:sing}, $G_\Omega(z)=\vartheta_3(-iz/2,e^{-1/2})$, whose zeroes sit at $z=-\tfrac12-n+i\pi(1+2m)$, $m,n\in\mathbb Z$ which is a single line $\phi=\pi$, diametrically opposite the state's peak at $\phi=0$, populated at every half-integer radius $\ell\in\mathbb Z+\tfrac12$. For a state centered at $z_0=\ell_0-i\phi_0$ the zero line sits at $\phi=\phi_0+\pi$, $\ell\in\ell_0+\mathbb Z+\tfrac12$ \cite{PhysRevResearch.5.L032006}.

The mathematical structure of the wrapped coherent state closely resembles that of the \textit{ideal} GKP state, one way of defining qubits or qudits through discretization of  continuous variables \cite{PhysRevA.97.032346}:
\begin{equation}\label{GKP}
\ket{\text{GKP}}=e^{-\Delta \hat{n} } \ket{\text{GKP}^{\text{ideal}}} \ , \ \ket{\text{GKP}^{\text{ideal}}} = \sum_{\alpha \in L(\mu)} e^{-i\alpha_{1}\alpha_{2}}\ket{\alpha} \, ,
\end{equation}
now in two dimensions, where $\hat n$ is the Fock number operator $\hat{n}\ket{n}=n\ket{n}$, playing here the role taken by $\hat L$; $\ket\alpha$ is the coherent state of $h(3)$, and $L(\mu)$ the state lattice of the code labelled by $\mu$, with $\Delta$ the width of each peak. Equations~\eqref{wrappednumber} and \eqref{GKP} share the same structure, an exponential of the number operator applied to a fiducial lattice state, and the Husimi distribution of the GKP state likewise has infinitely many zeroes~\cite{Chabaud:2020th}, reinforcing the correspondence. This is not a coincidence: we have already noted that, from a quantum-computational standpoint, the OAM eigenstate $\ket\ell$ plays a role analogous to the coherent state $\ket\alpha$, so that a superposition of OAM eigenstates is correspondingly analogous to a superposition of coherent states. Likewise, a coherent superposition of two wrapped coherent states with different amplitudes retains an infinite number of zeroes~\cite{PhysRevResearch.5.L032006}, mirroring a superposition of GKP states with different displacements.

\section{Distinguishing OAM states with the same numbers of zeroes}\label{OAMsecdis}

\subsection{Definition of the $L^{r}$ norm, and Wehrl entropy}
In this section we discuss several metrics for quantifying localization in phase space. The $L^{r}$ norm of a function $f(x)$ is defined by $\|f\|_{r} = \left( \int |f(x)|^{r} \, dx \right)^{1/r}$ for any $r \ge 1$ (or, for a discrete function or vector, by the corresponding sum). The $L^{r}$ norm quantifies the magnitude of $f(x)$ in a way that increasingly emphasizes its largest values as $r$ grows, making it increasingly sensitive to the peak of the distribution rather than to its tails. The set of functions with finite $L^{r}$ norm forms an $L^{r}$ space, a Banach space namely, a complete normed vector space. When $f$ is the Husimi distribution, the $L^{r}$ norm has been studied on the plane and the torus~\cite{Prosen1995QSOS,nonnenmacher_chaotic_2006} as an invariant fluctuation measure to characterize and distinguish quantum eigenstates. For $r=2$ it serves as a measure of quantum localization and has been used on the sphere as an indicator of quantumness~\cite{Goldberg:2020aa}.
\begin{figure*}
  \centering
  \begin{subfigure}[b]{0.18\textwidth}
    \centering
    \includegraphics[width=\textwidth]{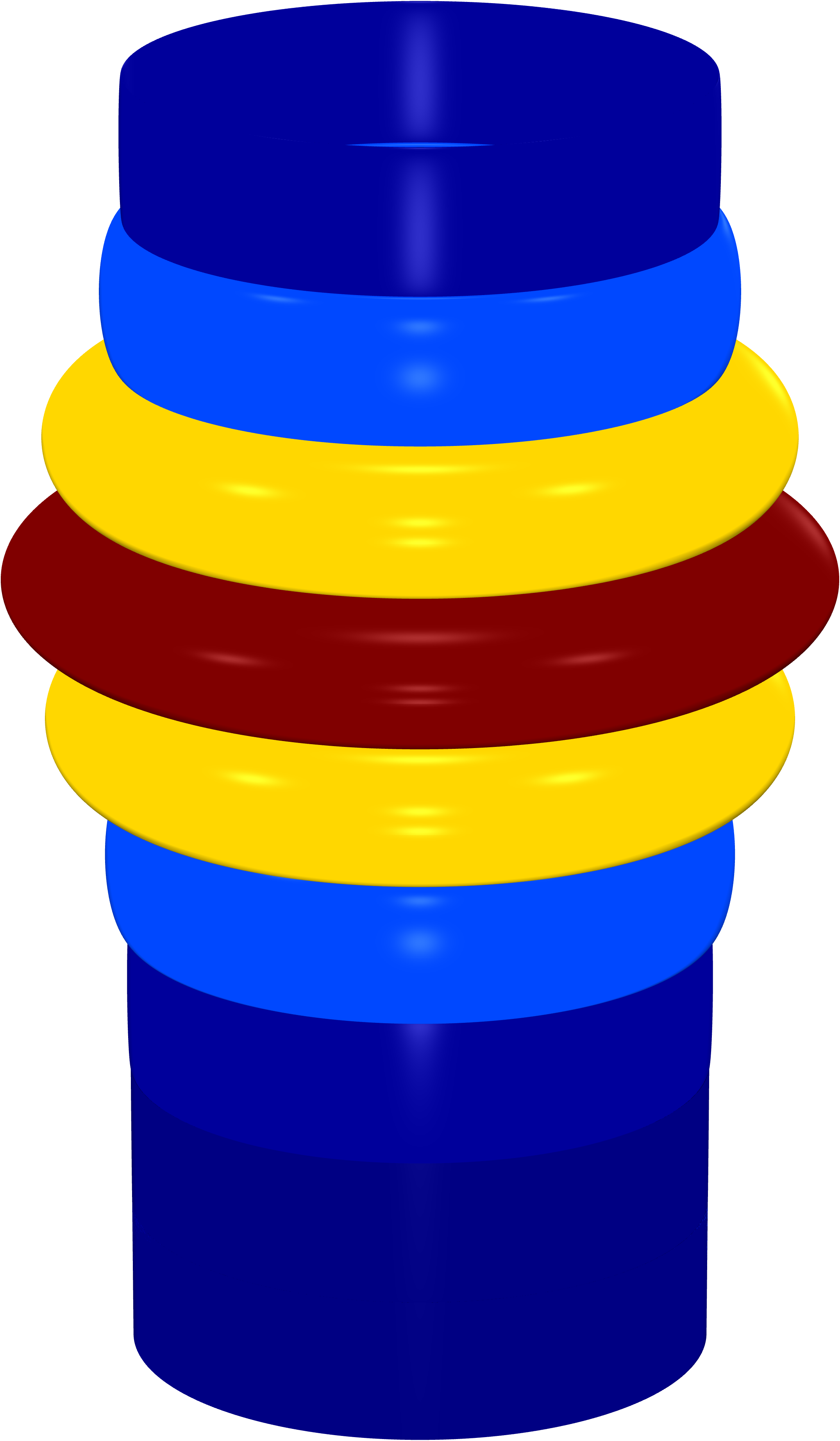}
    \caption{}
    \label{fig:hierarchy_a}
  \end{subfigure}
  \hfill
  \begin{subfigure}[b]{0.18\textwidth}
    \centering
    \includegraphics[width=\textwidth]{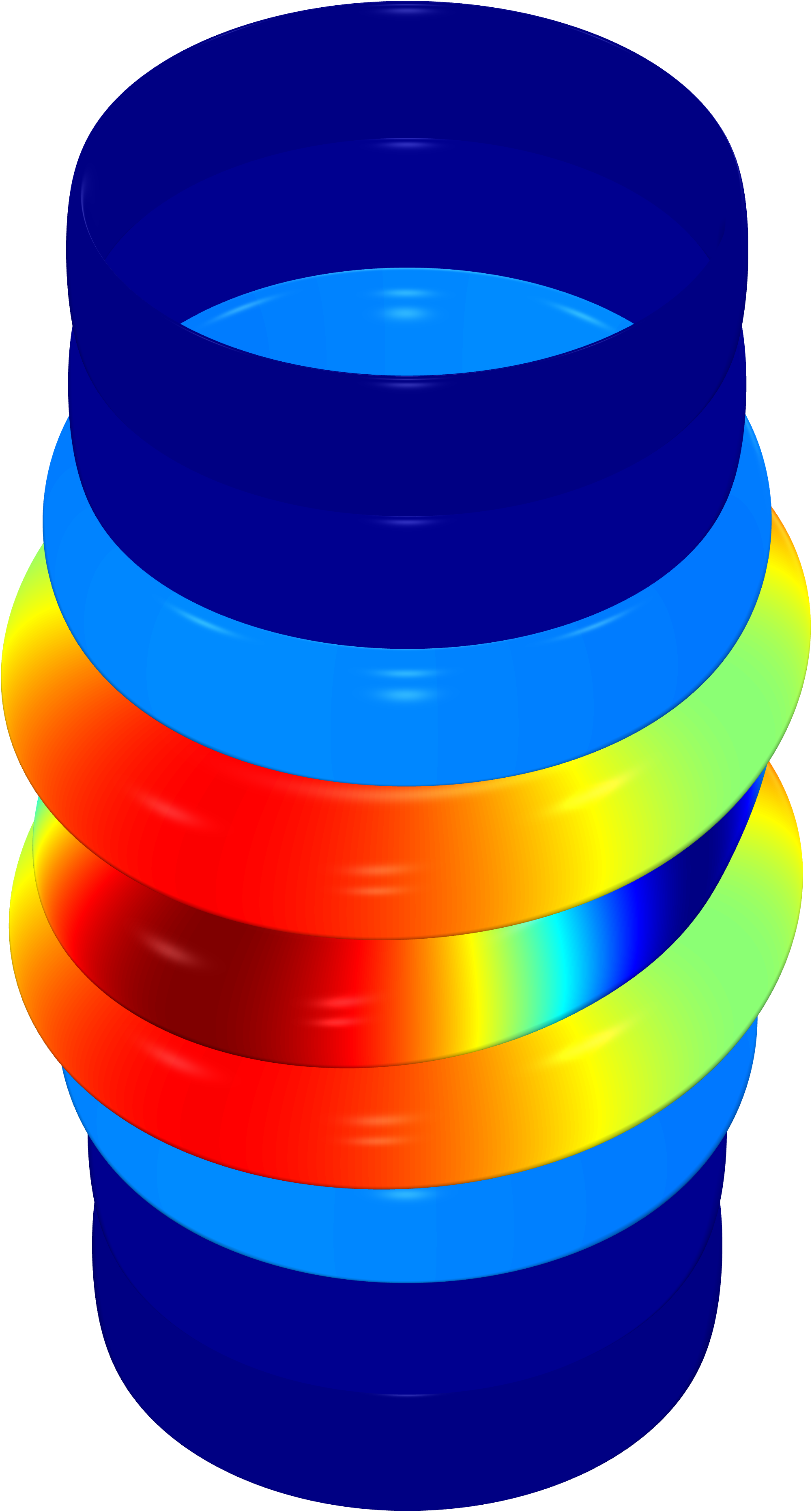}
    \caption{}
    \label{fig:hierarchy_b}
  \end{subfigure}
  \hfill
  \begin{subfigure}[b]{0.18\textwidth}
    \centering
    \includegraphics[width=\textwidth]{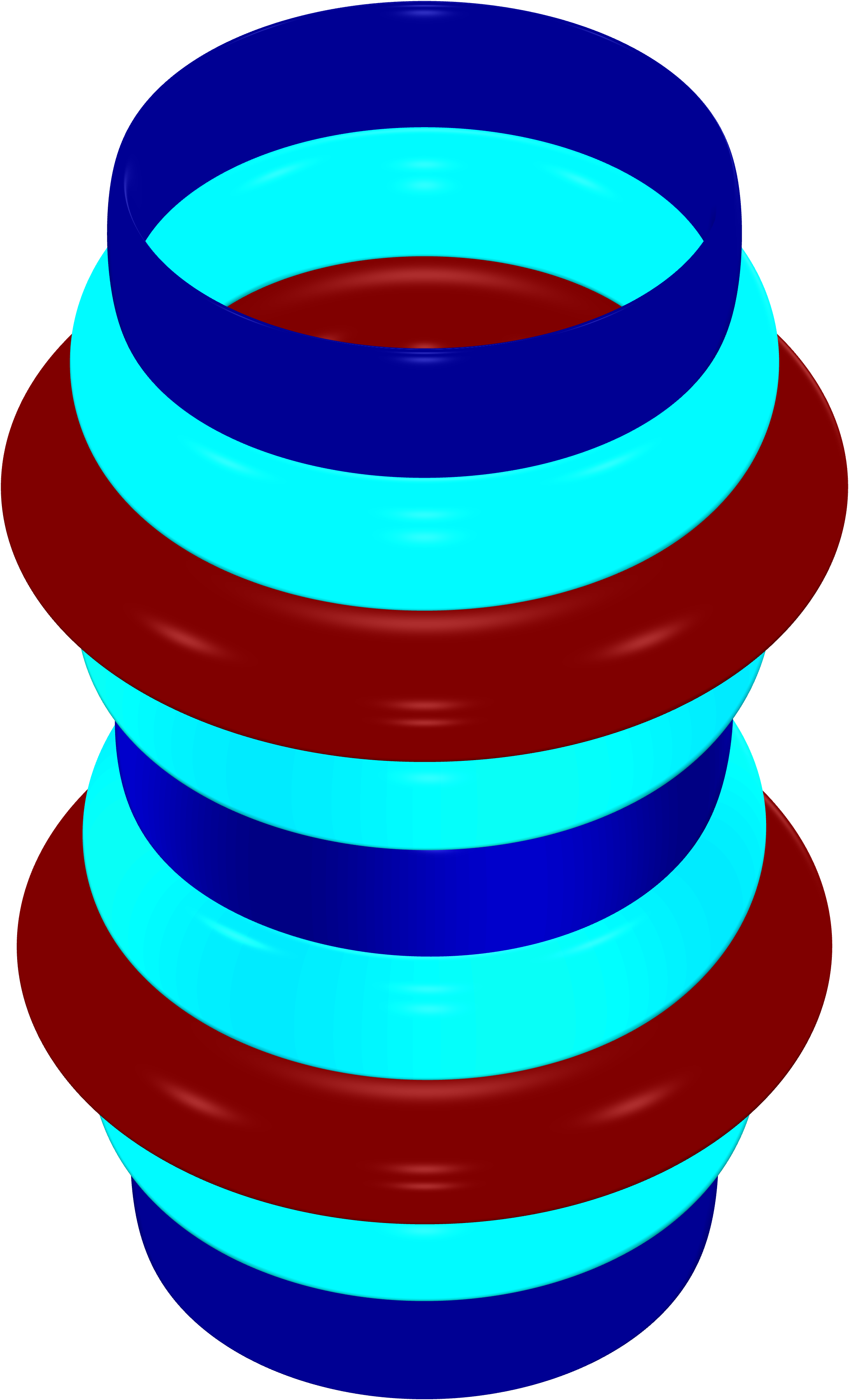}
    \caption{}
    \label{fig:hierarchy_c}
  \end{subfigure}
  \hfill
  \begin{subfigure}[b]{0.18\textwidth}
    \centering
    \includegraphics[width=\textwidth]{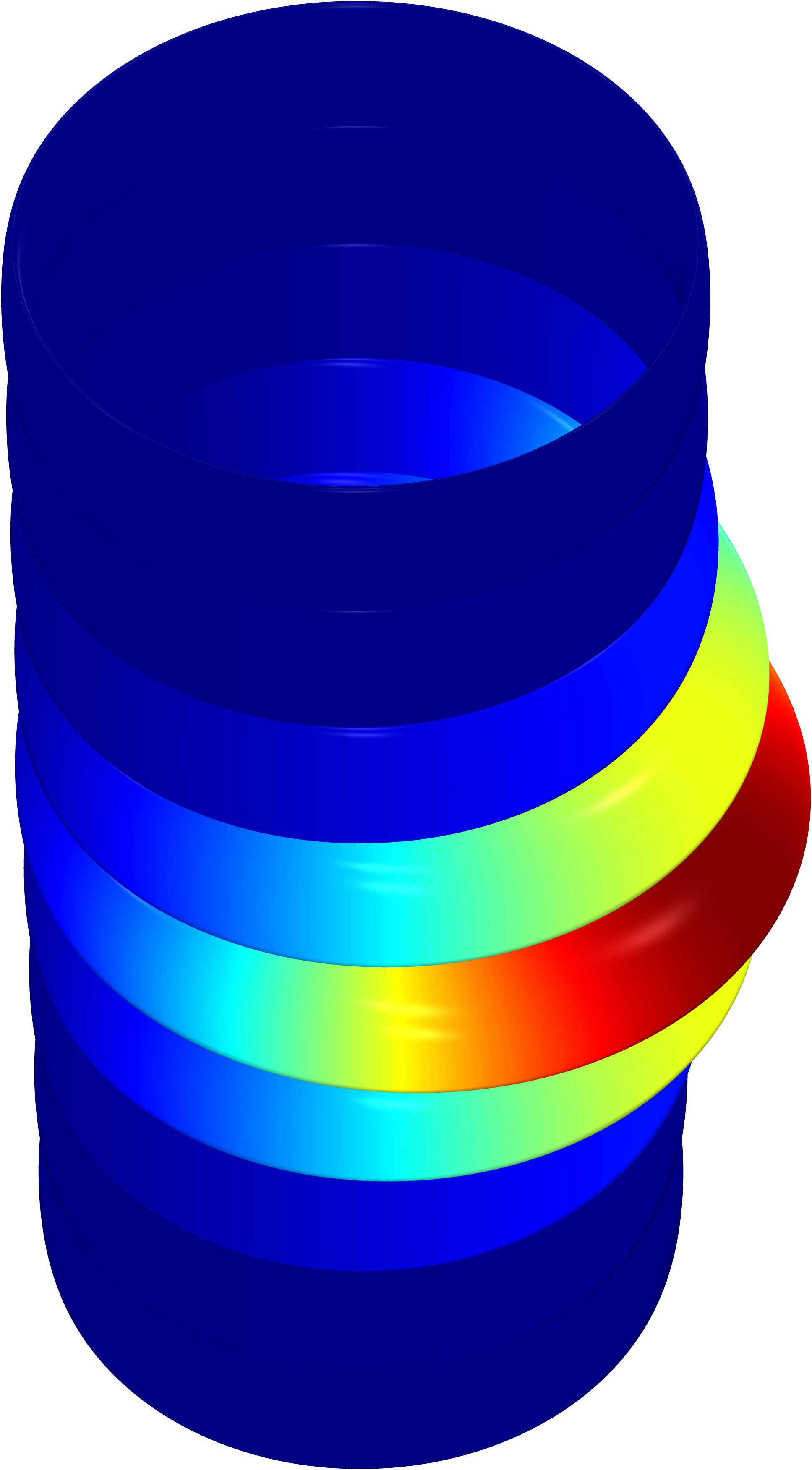}
    \caption{}
    \label{fig:hierarchy_d}
  \end{subfigure}
  \hfill
  \begin{subfigure}[b]{0.18\textwidth}
    \centering
    \includegraphics[width=\textwidth]{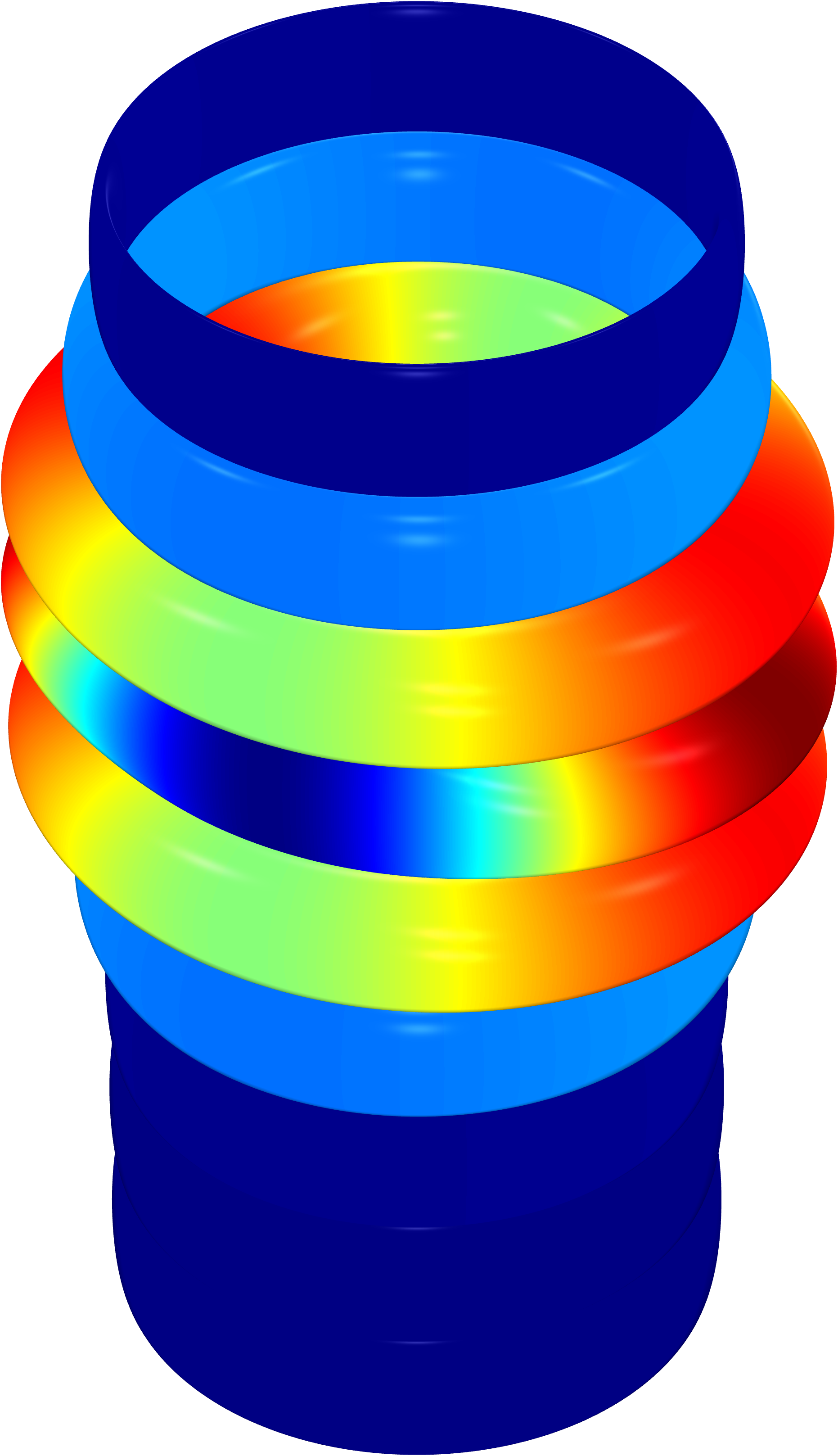}
    \caption{}
    \label{fig:hierarchy_e}
  \end{subfigure}

  \vspace{0.3cm}

  \begin{subfigure}[b]{0.18\textwidth}
    \centering
    \includegraphics[width=\textwidth]{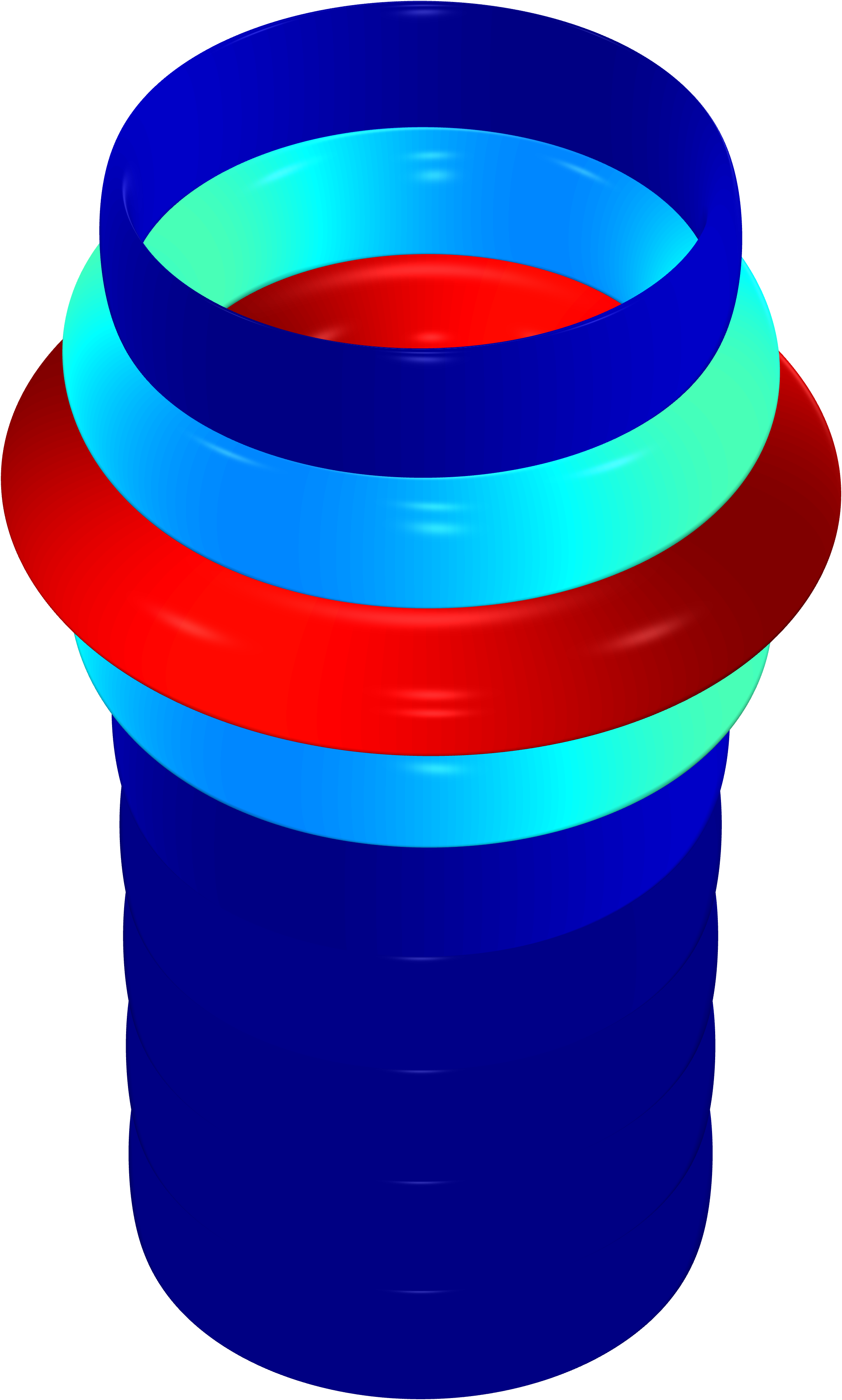}
    \caption{}
    \label{fig:hierarchy_f}
  \end{subfigure}
  \hfill
  \begin{subfigure}[b]{0.18\textwidth}
    \centering
    \includegraphics[width=\textwidth]{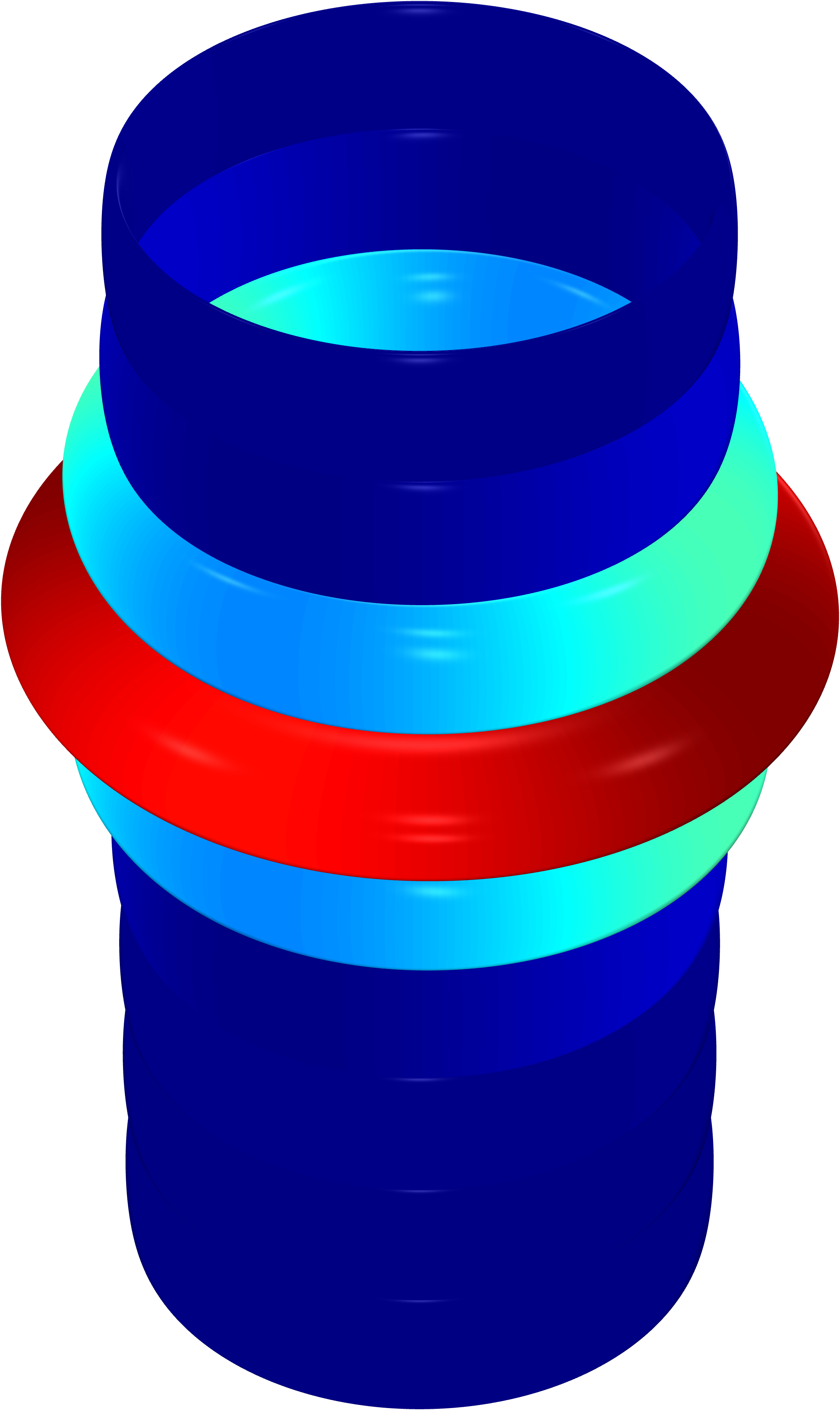}
    \caption{}
    \label{fig:hierarchy_g}
  \end{subfigure}
  \hfill
  \begin{subfigure}[b]{0.18\textwidth}
    \centering
    \includegraphics[width=\textwidth]{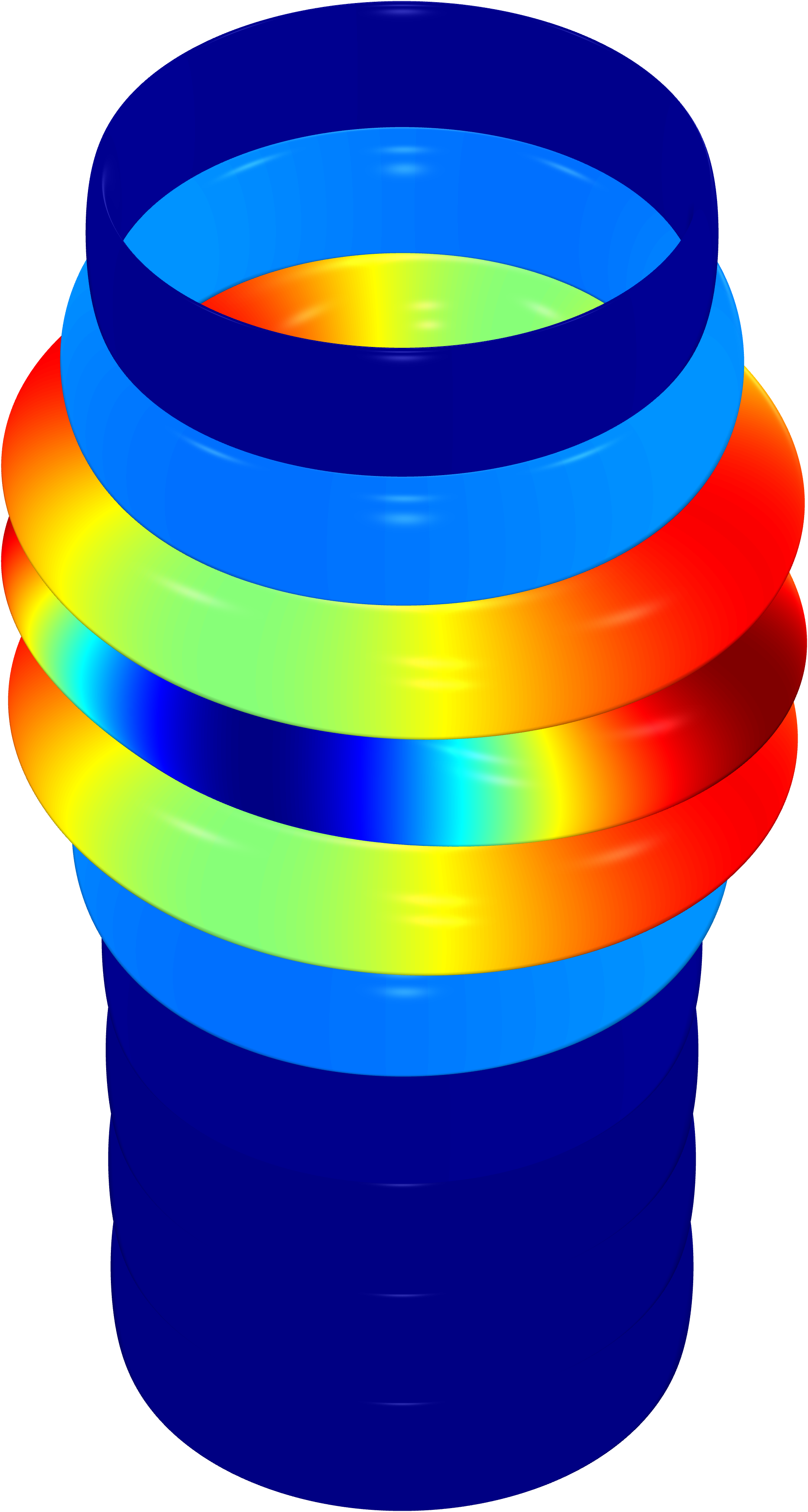}
    \caption{}
    \label{fig:hierarchy_h}
  \end{subfigure}
  \hfill
  \begin{subfigure}[b]{0.18\textwidth}
    \centering
    \includegraphics[width=\textwidth]{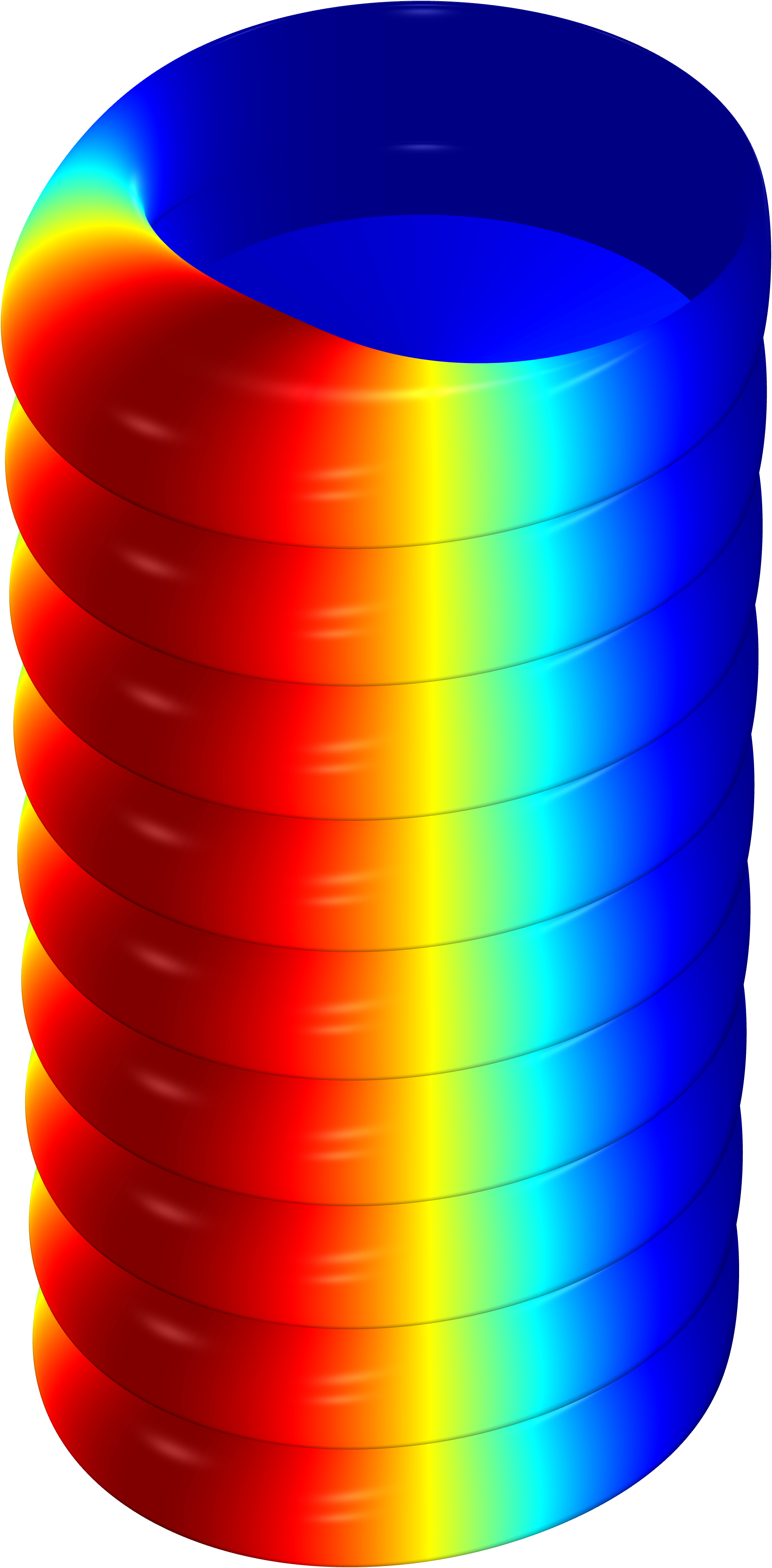}
    \caption{}
    \label{fig:hierarchy_i}
  \end{subfigure}
  \hfill
  \begin{subfigure}[b]{0.18\textwidth}
    \centering
    \includegraphics[width=\textwidth]{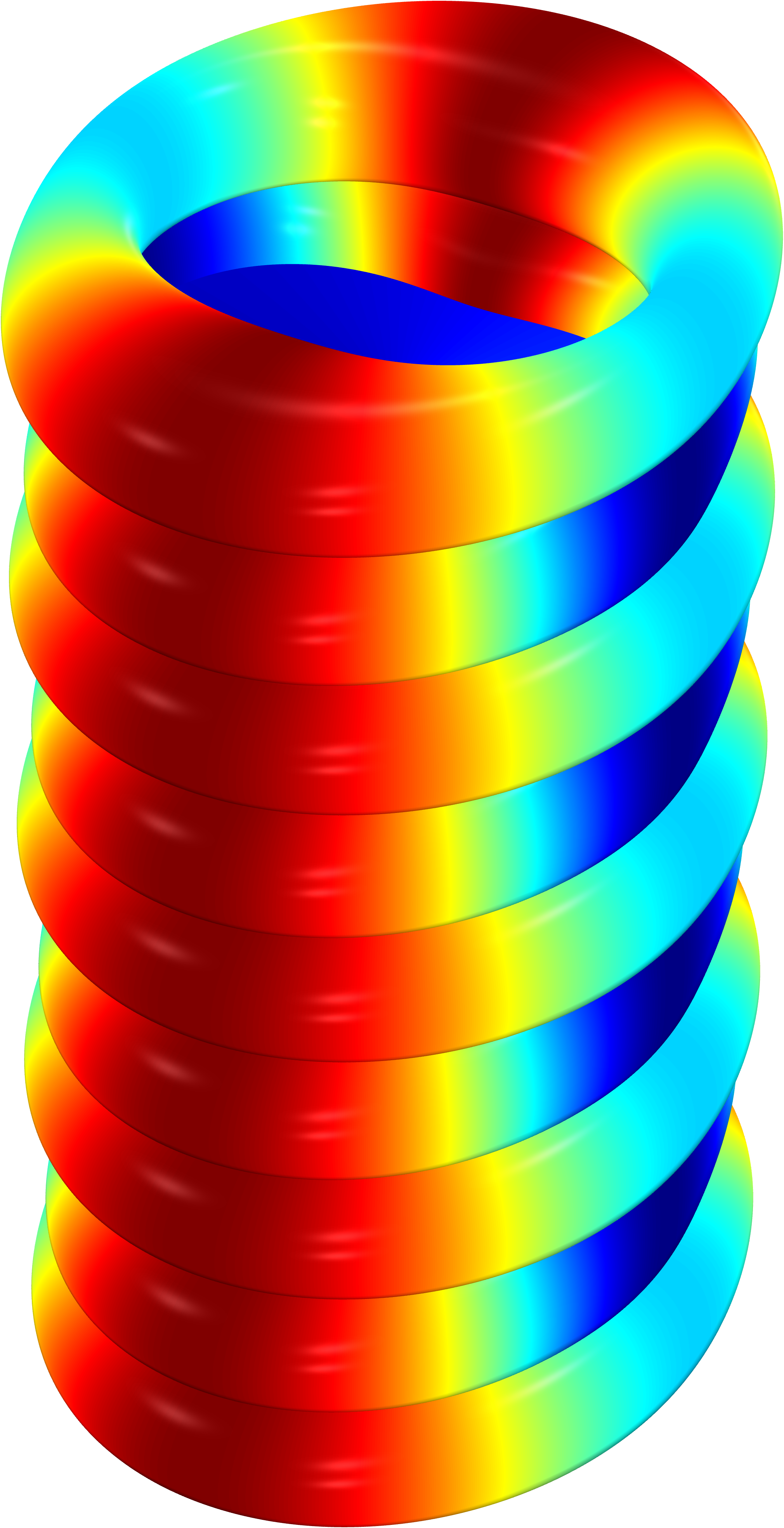}
    \caption{}
    \label{fig:hierarchy_j}
  \end{subfigure}

  \caption{Husimi distribution on the discrete cylinder. Each state is rendered as a stack of tori wrapped around a common cylindrical axis: the axis itself represents $\ell$, the azimuthal angle represents $\phi$, and at every height the tube's radius is set proportional to the local value of $Q_\psi(\ell,\phi)$ and colored on the same scale, from dark blue at the smallest values to red at the largest. Regions where $Q_\psi$ is negligible remain flush with the uniform, dark-blue background cylinder, so that the bulges and color bands trace out the shape, location, and zeroes of the Husimi distribution directly on the cylinder's surface. One caveat of this direct, linearly normalized rendering is worth noting: for two-peaked OAM superpositions, whose global amplitude is dominated by the peaks at $\ell=\pm\ell_0$ while the interference fringes sit at $\ell=0$. A linear mapping to torus radius strongly suppresses the radial signature of the fringes, whose amplitude is much smaller than that of the peaks; the interference nonetheless remains visible in the color scale even where the radial modulation is barely perceptible. (a) Husimi distribution of OAM eigenstate $l_{0}=1$. (b) Superposition of OAM eigenstate $l_{0}=1$. (c) Superposition of OAM eigenstate $l_{0}=2$. (d) Wrapped coherent state $(l_{0},\phi_{0})=(0,0)$. (e) Even cat state for $l_{0}=1$. (f) Even cat state for $l_{0}=2$. (g) Odd cat state for $l_{0}=1$, which is very closed to (f). (h) Odd cat state for $l_{0}=2$, which is very closed visually to (g). (i) Phase eigenstate for $\phi_{0}=\pi/4$. (j) Superposition of phase eigenstate  with $\phi_{0}=0,\pi$.}
  \label{husimi}
\end{figure*}
Here we extend the $L^{r}$ norm to the Husimi distribution on the discrete cylinder,
\begin{equation}
    \| Q_{\psi} \|_{r} = \left[\int_{0}^{2\pi}\frac{d\phi}{2\pi}\sum_{\ell\in\mathbb{Z}} \big[Q_{\psi}(\ell,\phi)\big]^{r}\right]^{1/r},
\end{equation}
with $r>0$. Because the state is normalized, $\|Q_{\psi}\|_{1}$ is always equal to one. As we will see, it serves as a measure of the localization and density of quantum states in phase space, though it can fail to distinguish states that are very different in nature, and it does not measure how spread the zeroes are.

$\|Q_\psi\|_{2}$ is quartic in the wavefunction, much like the inverse participation ratio (IPR), a common measure of quantum localization,
\begin{equation}
\text{IPR} = \frac{1}{\sum_{i} p_{i}^2}, \qquad p_i = |\psi_i|^2.
\end{equation}
Since $\sum_i p_i=1$, the IPR is always at least one, reaching its minimum, $\text{IPR}=1$, for a state fully localized on a single basis state. A lower IPR therefore indicates stronger localization and a higher IPR greater delocalization -- for instance, $\text{IPR}=1$ for an OAM eigenstate, while it is much larger for the (infinitely spread) wrapped coherent state. The choice of $r$ in the $L^{r}$ norm allows tuning the sensitivity of the metric to different features of the distribution, with $r=2$ being particularly natural due to its connection to Hilbert space structure. This measure is especially useful in studies of Anderson localization, many-body localization~\cite{misguich_inverse_2016}, quantum chaos~\cite{flambaum_entropy_2001,lakshminarayan_entanglement_2016}, and topological phase transitions~\cite{calixto_inverse_2015}, as it provides a single number that captures the extent to which a state occupies many or just a few sites (or basis elements). If the state is well spread over phase space, $Q_{\psi}(\ell,\phi)$ is nearly uniform and $\|Q_{\psi}\|_{2}^{2}$ is relatively small; if instead the state is concentrated in a small region, $Q_{\psi}(\ell,\phi)$ is sharply peaked and $\|Q_{\psi}\|_{2}^{2}$ is correspondingly larger.

To capture localization directly in phase space, we replace the coefficients $p_i^2$ in the IPR with the coefficients of the multipole expansion of the Husimi distribution on the discrete cylinder~\cite{PhysRevResearch.5.L032006}. The multipole expansion decomposes the Husimi distribution into angular and radial components, with coefficients $q_{nk}$ capturing the contributions of the different multipole moments, yielding a phase-space IPR with the same physical interpretation:
\begin{equation}
    \text{IPR} = \frac{1}{\sum_{n,k\in\mathbb{Z}^{2}} |q_{nk}|^2} \ ,  \ q_{nk} = \frac{1}{2\pi} \int_{0}^{2\pi} Q_{\psi}(k,\phi)\, e^{in\phi} \, d\phi,
\end{equation}
again giving a localization measure quartic in the wavefunction. The Wehrl entropy~\cite{RevModPhys.50.221,floerchinger_wehrl_2021}, defined from the Husimi distribution, generalizes to the discrete cylinder as
\begin{equation}
S_{W}=-\int_{0}^{2\pi} \frac{d\phi}{2\pi}\sum_{\ell\in \mathbb{Z}} Q_{\psi}(\ell,\phi) \ln Q_{\psi}(\ell,\phi).
\end{equation}

The higher the $L^{r}$ norm at large $r$, the more localized the state. Conversely, the higher the IPR, the more delocalized the state, as follows directly from its definition. The Wehrl entropy tracks phase-space localization in the same way as the IPR, as we shall see.

\subsection{OAM eigenstate and superposition}
\begin{figure*}
  \centering
  \begin{subfigure}[b]{0.4\textwidth}
    \centering
    \includegraphics[width=\textwidth]{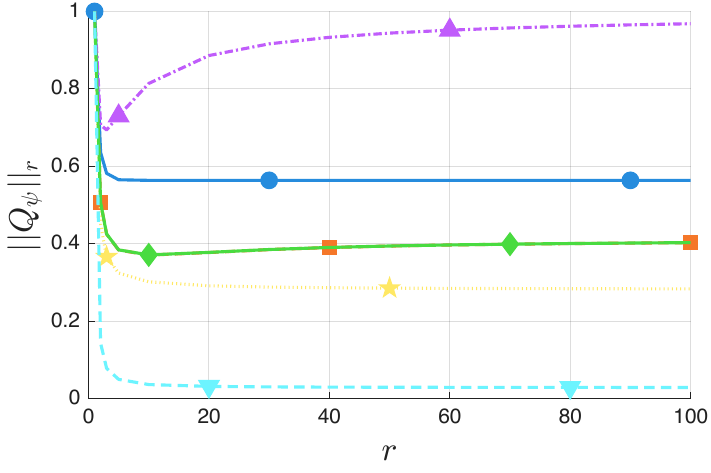}
    \caption{}
    \label{fig:OAM_a}
  \end{subfigure}
  \hfill
  \begin{subfigure}[b]{0.4\textwidth}
    \centering
    \includegraphics[width=\textwidth]{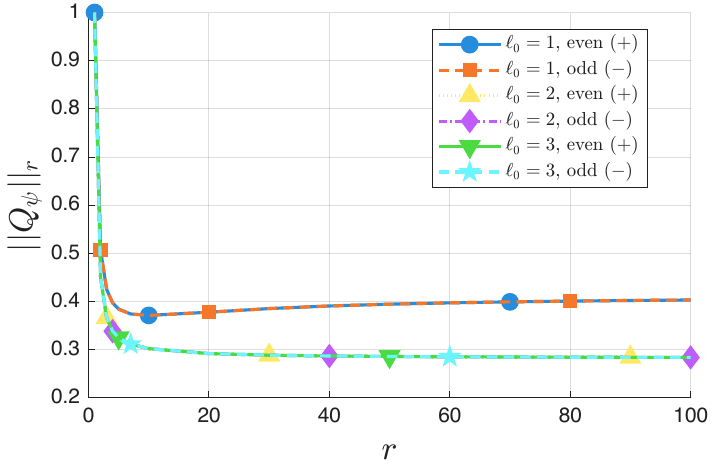}
    \caption{}
    \label{fig:OAM_b}
  \end{subfigure}

  \vspace{0.3cm}

  \begin{subfigure}[b]{0.4\textwidth}
    \centering
    \includegraphics[width=\textwidth]{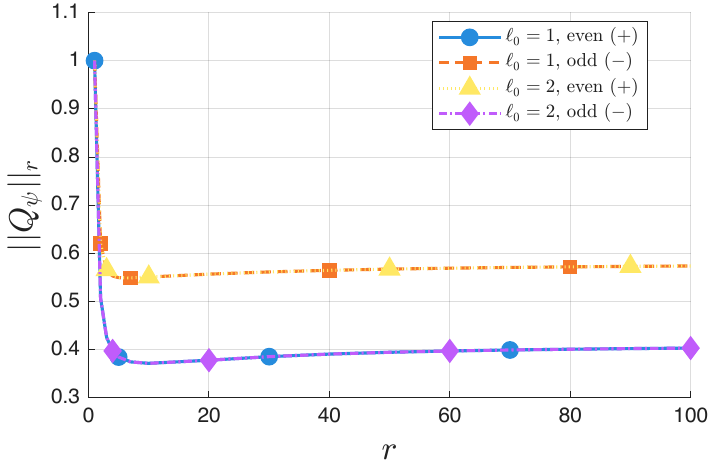}
    \caption{}
    \label{fig:OAM_c}
  \end{subfigure}
  \hfill
  \begin{subfigure}[b]{0.4\textwidth}
    \centering
    \includegraphics[width=\textwidth]{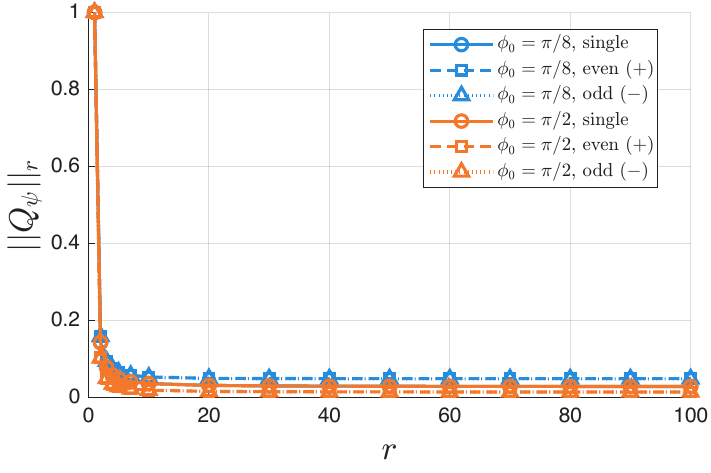}
    \caption{}
    \label{fig:OAM_d}
  \end{subfigure}
  \caption{(a)$L^{r}$ norm for the OAM eigenstate (circle), its superposition for $\ell_{0}=1$ (square) and $\ell_{0}=2$ (star), the coherent state (upward-pointing triangle), the even cat state with $\ell_{0}=2$ and lobes at $\phi_{0}=0,\pi$ (rhombus), and the phase eigenstate (downward-pointing triangle). The superposition with $\ell_{0}=1$ has the same norm as the even cat state at even $\ell_{0}$, since both states have similar Husimi distributions (see Fig.~\ref{husimi}(b) and (e); this measure alone therefore cannot distinguish between the two. (b) $L^{r}$ norm for the even ($+$) superposition of OAM eigenstates, for several values of $\ell_{0}$, both odd and even. (c) $L^{r}$ norm for the even and odd cat-state superpositions, with $\phi_{0}$ fixed so that the two lobes sit at $\phi=0$ and $\phi=\pi$. (d) $L^{r}$ norm for the phase eigenstate.}
  \label{OAMLr}
\end{figure*}

The Husimi distribution of the OAM eigenstate~\cite{PhysRevResearch.5.L032006} is
\begin{equation}
Q_\psi(\ell,\phi) = \frac{1}{\varpi}\, e^{-(\ell-\ell_{0})^{2}},
\end{equation}
shown in Fig.~\ref{fig:hierarchy_a}: a central $\phi$-independent ring with no zero. For the OAM eigenstate $\ket{\ell_{0}}$ the $L^{r}$ norm can be computed in closed form,
\begin{equation}
    \|Q\|_{r}^{r} = \frac{1}{\varpi^{r}} \sum_{\ell\in\mathbb{Z}} e^{-r(\ell-\ell_{0})^{2}} = \frac{1}{\varpi^{r}} \vartheta_{3}(0, e^{-r}),
\end{equation}
and since this expression does not depend on $\ell_{0}$, the eigenstate's degree of localization is the same for every $\ell_{0}$, as confirmed in Fig.~\ref{OAMLr}.

The stellar representation of a superposition of OAM states, $\ket{\psi}_{\pm}=\frac{1}{\sqrt{2}}(\ket{\ell_{0}}\pm\ket{-\ell_{0}})$, can be written as
\begin{equation}
   F_{\psi_{\pm}}(\ell, \phi) = \frac{1}{\sqrt{2\varpi}} e^{-\ell^{2}/2} \left( e^{-\ell_{0}^{2}/2} e^{\ell_{0}(\ell+i\phi)} \pm e^{-\ell_{0}^{2}/2} e^{-\ell_{0}(\ell+i\phi)} \right),
\end{equation}
with the corresponding Husimi distribution, shown in Fig.~\ref{fig:hierarchy_b} and \ref{fig:hierarchy_c}, given by
\begin{align}
    Q_{\psi_{\pm}}(\ell, \phi) = \frac{1}{2\varpi} e^{-\ell^{2}} e^{-\ell_{0}^{2}} \left( 2\cosh(2\ell_{0}\ell) \pm 2\cos(2\ell_{0}\phi) \right).
\end{align}
Its zeroes are located at $\ell=0$ and $\phi=(2k+1)\pi/(2\ell_{0})$, for $k=1,\dots,2\abs{\ell_{0}}-1$; the number of zeroes therefore increases by two for each unit increase of $\ell_{0}$, starting from $\ell_{0}=1$. The Husimi distribution consists of two such rings, at $\ell=\pm\ell_{0}$, with an interference pattern between them visible only for $\ell_{0}=1$ (see Fig.~\ref{husimi}(b),(c)).

\begin{figure*}
  \centering
  \begin{subfigure}[b]{0.99\textwidth}
    \centering
    \includegraphics[width=\textwidth]{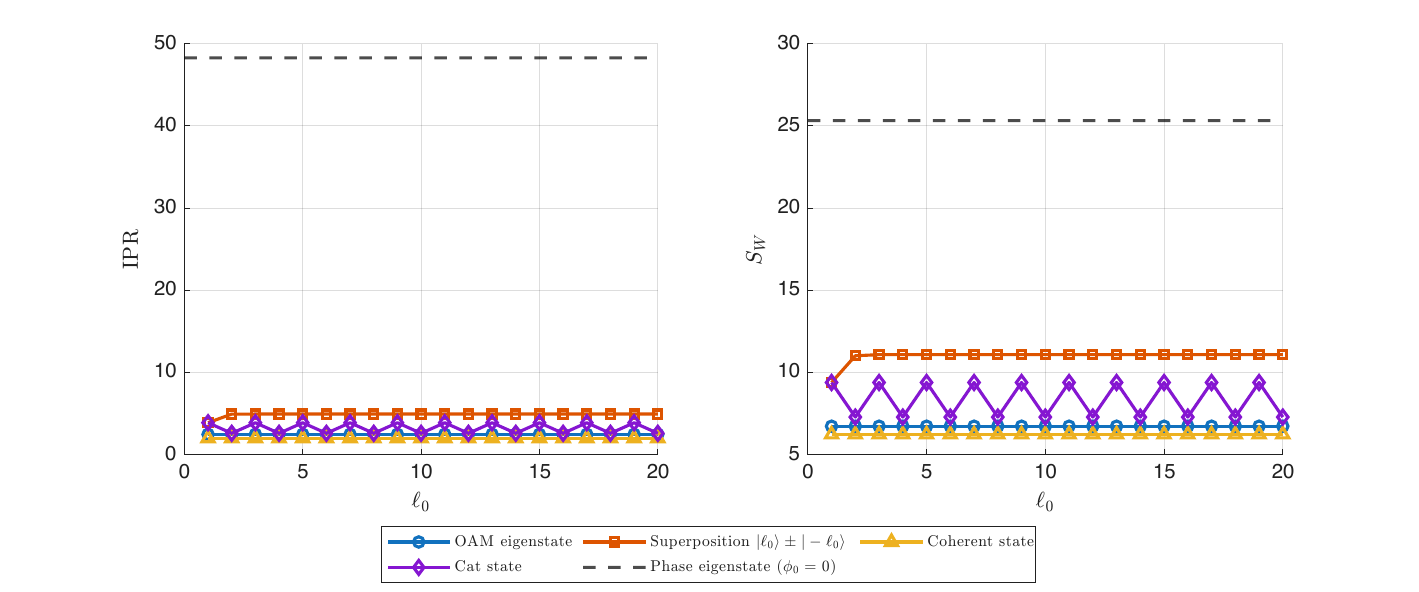}
    \caption{}
    \label{fig:IPR_a}
  \end{subfigure}
  \begin{subfigure}[b]{0.99\textwidth}
    \centering
    \includegraphics[width=\textwidth]{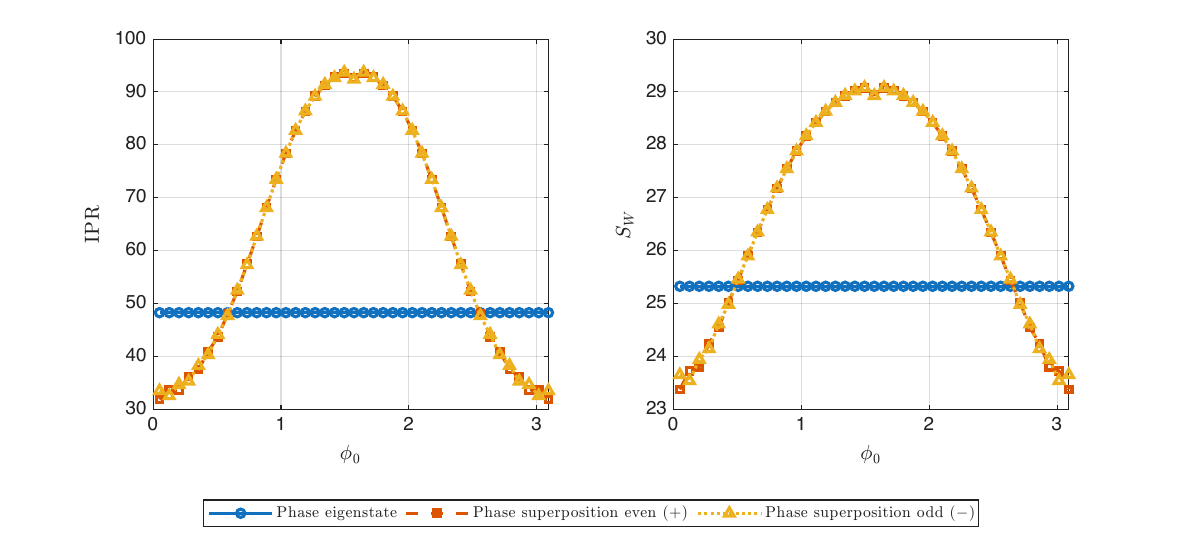}
    \caption{}
    \label{fig:IPR_b}
  \end{subfigure}
  \caption{(a) IPR and Wehrl entropy for the OAM eigenstate, its superposition, the coherent state, and the even cat state. We can clearly see the different between the even and odd value of $l_{0}$ in terms of delocalisability, which is also clearly visible in Fig.~\ref{fig:hierarchy_e} and \ref{fig:hierarchy_f}. (b) IPR and Wehrl entropy for the phase eigenstate and its superposition. The higher the IPR or Wehrl entropy, the more delocalized the state is in phase space.}
  \label{fig:IPR}
\end{figure*}

The $\|Q\|_{r}$ norm of the OAM superposition is shown in Fig.~\ref{OAMLr}, and is lower than that of a single OAM eigenstate, indicating a more delocalized state. The superposition with $\ell_{0}=1$ has a larger $L^{r}$ norm than that with $\ell_{0}=2$: once the two components are sufficiently separated, they no longer overlap, and the interference term is damped, so that the Husimi distribution becomes two separate rings with no visible interference between them. The IPR, and similarly the Wehrl entropy (see Fig.~\ref{fig:IPR}(a)), show the same behavior: the $\ell_{0}=1$ superposition can be distinguished from those with $\ell_{0}\ge2$, which are essentially indistinguishable from one another. The IPR and the Wehrl entropy are both higher for the OAM superposition than for the other states considered, again confirming that it is the more delocalized state.

\subsection{Wrapped coherent state and cat-state superpositions}

We now consider the coherent state $\ket{\ell_{0}, \phi_{0}}$, whose stellar representation is \cite{PhysRevResearch.5.L032006}
\begin{equation}
 F_{\psi}(l,\phi) = \frac{1}{\varpi}    e^{-\case{1}{2} (\ell^{2} + \ell_{0}^{2})} \,  \vartheta_{3} \left (\tfrac{1}{2} (\phi_{0} - \phi) + \tfrac{1}{2} i (\ell + \ell_{0})\big | e^{-1} \right ) \,  \, .
 \end{equation}
and whose Husimi distribution is represented in Fig.~\ref{fig:hierarchy_d} is instead peaked at $\phi=0$ but carries infinitely many zeroes running the length of the cylinder on the diametrically opposite side, $\phi=\pi$ which is the clearest visual signature of its place at the top of the hierarchy. We point out that the norm is substantially larger for the wrapped coherent state (see Fig.~\ref{OAMLr}), which can therefore be used as an indicator that the wrapped coherent state is more delocalized in phase space.  We observe that the IPR is higher for the superposition state than for the OAM eigenstate and the coherent state. This indicates that the IPR is not a measure of how delocalized the zeros are, but rather of how delocalized the state itself is: the coherent state is localized at $(\ell_0,\phi_{0})$, and although it has an infinite number of zeroes, they are strongly damped, whereas the superposition state has only a finite number of zeroes yet a more localized probability density.\\

The stellar representation of the odd and even cat states $\ket{\psi}_{\text{cat}\pm} = (\ket{\ell_{0}, \phi_{0}} \pm \ket{\ell_{0}, \phi_{0} + \pi})/\sqrt{\mathcal{N}_{0}}$ reads \cite{PhysRevResearch.5.L032006}
\begin{align}
F_{\mathrm{cat} \pm} (\ell, \phi)  = \frac{2}{\varpi\sqrt{  \mathcal{N}_{0}}}e^{-\case{1}{2} (\ell^{2}+\ell_{0}^{2})} (\vartheta_{3}( \tfrac{1}{2} (\phi_{0} - \phi) \\
+ \tfrac{1}{2} i (\ell + \ell_{0})\big | e^{-1} )\pm\vartheta_{3}  ( \tfrac{1}{2} (\phi_{0}+\pi - \phi) + \tfrac{1}{2} i (\ell + \ell_{0})\big | e^{-1}  )).
\end{align}
whose Husimi distribution is represented in Fig.~\ref{fig:hierarchy_e}, \ref{fig:hierarchy_f}, \ref{fig:hierarchy_g}, \ref{fig:hierarchy_h} for $l_{0}=1,2$. $\mathcal{N}_{0}$ is a normalisation constant that will be calculated numerically. Figures~\ref{husimi}(e)--(h) show the even and odd superpositions of the wrapped coherent state for $\ell_0=1,2$; all four carry the same four zeroes, yet their apparent spread differs with the parity of $\ell_0$, the even superposition at odd $\ell_0$ appearing as delocalized as the odd superposition at even $\ell_0$, and conversely, the visual counterpart of the localization measures of Sec.~\ref{OAMsecdis}.

 Finally, the $L^{r}$ norm of the wrapped coherent state is larger than that of the cat states for fixed $l_{0}$ and $\phi_{0}$, which again reflects the fact that the wrapped coherent state is more localized on the discrete cylinder. As the OAM eigenstate does not have any zero, and as a higher $L^{r}$ norm, this ensures that a high $L^{r}$ norm does not indicate a number of zeroes delocalisability, but density localisability. This is therefore the exact opposite of the planar case, where an infinitely squeezed cat state can have an infinite number of zeroes.

\subsection{Phase eigenstate and its superposition}
The stellar representation of the phase eigenstate $\ket{\phi_{0}}$ is
\begin{equation}
    F_{\phi_{0}}(\ell,\phi)=\frac{1}{\sqrt{2\pi \varpi}} e^{i\ell(\phi-\phi_{0})} \vartheta_{3}\!\left(\frac{1}{2}(\phi-\phi_{0})\,\middle|\,e^{-1/2}\right),
\end{equation}
whose associated Husimi distribution, shown in Fig.~\ref{fig:hierarchy_i} for $\phi_{0}=\pi/4$, confirms that the state is sharply localized at $\phi=\phi_{0}$ but fully delocalized along $\ell$, and carries no zero. Its $L^{r}$ norm, shown in Fig.~\ref{OAMLr}, is the lowest among all the states considered.

The stellar representation for the superposition of phase eigenstates $\ket{\psi}=\frac{1}{\sqrt{2}}(\ket{\phi_{0}}+\ket{-\phi_{0}})$ is then
\begin{equation}
\begin{split}
F_{\psi}(\ell,\phi)
&=\frac{1}{\sqrt{4\pi \varpi}}e^{i\ell\phi}
\Bigl(
e^{-i\ell\phi_{0}}
\vartheta_{3}\!\left(
\frac{\phi-\phi_{0}}{2}\,\middle|\,e^{-1/2}
\right) \\
&\qquad\qquad
+e^{i\ell\phi_{0}}
\vartheta_{3}\!\left(
\frac{\phi+\phi_{0}}{2}\,\middle|\,e^{-1/2}
\right)
\Bigr),
\end{split}
\end{equation}
whose associated Husimi distribution, shown in Fig.~\ref{fig:hierarchy_j} for $\phi_{0}=\pi/4$, exhibits infinitely many oscillations, and zeros, along two heights of the cylinder which is the clearest signature, among all the states considered, of a highly quantum character. The superposition of phase eigenstates clearly has zeroes and is delocalized in phase space; its $L^{r}$ norm is lower than that of the other states, consistent with the observation above that a higher $L^{r}$ norm indicates a more localized state.

Although the phase eigenstate has no zero, it is the most delocalized of all the states considered, and accordingly its IPR and Wehrl entropy are higher than those of the others. The superposition of phase eigenstates is even more delocalized than the phase eigenstate itself, as indicated by the IPR and Wehrl entropy (see Fig.~\ref{fig:IPR_b}); the delocalization is maximized when the two phase eigenstates are exactly opposed to each other.

\section{Conclusion}
We have established a hierarchy of stellar rank on the discrete cylinder, ranked by the number of zeroes of the stellar distribution based on quantum computing usefulness. OAM and phase eigenstates sit at its bottom: their associated stellar representations have no zero and a positive Wigner distribution, hence are classically simulable. The OAM eigenstates are invariant, up to a phase, under the natural translations of the cylinder, unlike Fock states under phase-space displacement; this is why every OAM eigenstate, and not only a distinguished vacuum, sits at the bottom of the hierarchy. Many twisted-photon states nonetheless share the same rank, the odd and even cat-state superpositions of the wrapped coherent state always carry four zeroes, independently of $\ell_{0}$, and must then be told apart by complementary quantities, such as the localization measures introduced above, or those of \cite{Chabaud:2020th}. A sharper, resource-theoretic reading of the hierarchy itself remains to be developed: qubit computation has its Clifford/non-Clifford distinction, and continuous-variable computation its Gaussian/non-Gaussian one, and it would be worth asking whether the discrete cylinder supports an analogous free-versus-resource split tied to the number of zeroes, one route being a mapping onto bosonic operators \cite{fan_cooper-pair_2006}.

Our choice of the coherent state on the discrete cylinder, the wrapped coherent state which sits at the top of the hierarchy, was motivated by its close mathematical resemblance to the GKP state. This resemblance may run deeper than analogy: using modular variables, the continuous plane and a double discrete cylinder have been shown to give equivalent phase-space representations~\cite{fabre_wigner_2020}. This suggests that the discrete cylinder, rather than the plane, may be the more natural phase-space geometry for the GKP encoding itself. The von Mises coherent state offers another candidate, defined instead by saturating the angle-OAM uncertainty relation rather than by a ladder-operator eigenvalue equation~\cite{pegg_minimum_2005,franke-arnold_uncertainty_2004,rehacek_experimental_2008,Mista:2022tv}. It, too, carries an infinite number of zeroes, confirming that the coherent states of the discrete cylinder are the most non-classical in the sense of this hierarchy.  Finally, it would be interesting to connect this hierarchy to metrological performance — for phase estimation with twisted photons in particular — and to the resource-theoretic picture sketched above.

\section*{ACKNOWLEDGMENT}
 N. Fabre acknowledges fruitful discussions with Ulysse Chabaud, Jack Davis,  Jean-Pierre Gazeau, Andrei Klimov, Romain Murenzi, and Luis Sanchez Soto for the completion of this manuscript. The author thank AI tool for assistance with English language editing.

\section*{Disclosures}

The authors declare no conflicts of interest.

\section*{Data availability}

Data underlying the results presented in this paper are not publicly available at this time but may be obtained from the authors upon reasonable request.

\onecolumngrid
\appendix

\section{Stellar representation into the discrete cylinder}\label{sec:bargmann}

\subsection{Covariant integral quantization on the cylinder}\label{app:covariant}
We remind the covariant integral quantization on the cylinder, already evoked in \cite{PhysRevResearch.5.L032006,Gazeau:2022aa}.
At a general level, constructing coherent states amounts to finding a map from $X$ to a closed subspace $\mathcal{K}\subset L^{2}(X,\mu)$, $X \ni x \mapsto |x \rangle \in \mathcal{K}$, satisfying \cite{PhysRevResearch.5.L032006,Gazeau:2022aa}
\begin{equation}
\label{condition}
\langle x| x \rangle = 1, \qquad \int_{X}d\mu(x) \; w(x) \ket{x}\bra{x}= \openone \, ,
\end{equation}
with $w(x)$ a weight function; the resulting family $\{\ket x\}$ is the set of coherent states sought~\cite{Gazeau:2009vb}. Let $\{\Upsilon_{n}(x)\}$ be an orthonormal set in $L^2(X,\mu)$ with $0<\mathcal M(x)\equiv\sum_{n\in\mathbb{Z}}|\Upsilon_n(x)|^2<\infty$, and let $\{\ket{e_n}\}$ be an orthonormal basis of a Hilbert space $\mathcal H$ -- not necessarily $\mathcal K$ -- in one-to-one correspondence with $\{\Upsilon_n(x)\}$. Then the states \cite{PhysRevResearch.5.L032006,Gazeau:2022aa}
\begin{equation}
\label{eq:CS}
\ket{x}  = \frac{1}{\sqrt{\mathcal{M}(x)}} \sum_{n\in\mathbb{Z}} \Upsilon_{n}^{\ast} (x) \ket{e_{n}}
\end{equation}
satisfy Eq.~\eqref{condition} with $w(x)= \mathcal{M}(x)$.  No choice of the functions $\Upsilon_n$ is preferred a priori; several candidates have been proposed~\cite{Gazeau:2022aa}, including the wrapped Gaussian~\cite{Ruzzi:2006dq}, the Dirichlet~\cite{Dirichlet:1829aa} and Fej\'er kernels~\cite{Fejer:1904aa}, and the von Mises distribution~\cite{pegg_minimum_2005,franke-arnold_uncertainty_2004,rehacek_experimental_2008,Mista:2022tv}. The natural framework is therefore the Hilbert space $L^{2}(X,\mu)$ of square-integrable functions on the phase space $X$ which is the discrete cylinder in our case with $\mu$ the associated invariant measure~\cite{PhysRevResearch.5.L032006,Gazeau:2022aa}. The cylinder phase space $X = \mathcal{S}_{1} \times \mathbb{Z}$ is parametrized naturally by $x \equiv (\ell, \phi)$, with measure $\int_{X}d\mu(x)= \frac{1}{2\pi}\sum_{\ell\in \mathbb{Z}}\int_0^{2\pi}d\phi$.

\subsection{Conformal transformation}\label{sec:conformal}

Different coordinate choices relate the discrete cylinder to the complex plane.  A convenient parametrisation is obtained by introducing a real longitudinal coordinate $l$ and an angular coordinate $\phi$ (periodic, $\phi\sim\phi+2\pi$), and defining
\begin{equation}
z = l - i\phi,\qquad \xi = e^{z} = e^{l}\,e^{-i\phi}.
\end{equation}
The map $\xi=e^{z}$ sends the cylinder onto the punctured complex $\xi$-plane $\mathbb{C}^{*}=\mathbb{C}\setminus\{0\}$: circles of constant $l$ become circles of radius $e^{l}$ centred at the origin, while lines of constant $\phi$ become radial rays.  Its inverse is $z=\log\xi$, with a branch cut that can be placed along any ray not intersecting the physical domain.

This exponential mapping is the standard tool to flatten a cylindrical geometry onto the plane, notably in radial quantisation where the Euclidean time coordinate on the cylinder becomes the logarithm of the radial coordinate \cite{Belavin:1984vu,DiFrancesco:1997nk,Ginsparg:1988ui}.  Even when the periodic direction is discretised---as is natural for the conjugate pair $(\phi,L_{z})$ on $\mathbb{S}^{1}\times\mathbb{Z}$---the conformal transformation retains the same exponential form, the discreteness being encoded in the $2\pi i$-periodicity of the resulting entire functions \cite{Kastrup:2016,Rigas:2017}.

Two singular points appear in the mapping: as $l\to-\infty$ the image $\xi\to0$, while $l\to+\infty$ corresponds to $|\xi|\to\infty$.  These are not artifacts of the discretisation but reflect the infinite extent of the cylinder.  Their physical impact is harmless and can be controlled by several standard strategies.  One may work directly on the punctured plane $\mathbb{C}^{*}$, or compactify by adding the point at infinity to obtain the Riemann sphere $\hat{\mathbb{C}}$.  Alternatively, a short-distance cutoff is introduced by excising a small disk $|\xi|<\epsilon$ (equivalently, restricting to $l>\log\epsilon$) and taking $\epsilon\to0$ after renormalisation of the relevant observables \cite{DiFrancesco:1997nk}.

\subsection{Analytic structure of the stellar function}\label{app:sing}

\subsubsection{Why $H_\psi(\xi)$, $\xi=e^{\ell-i\phi}$, is not entire}

Written as a function of $\xi$, the stellar function is the bilateral
Laurent series $H_\psi(\xi)=\sum_{n\in\mathbb Z}e^{-n^{2}/2}\psi_n\,
\xi^{n}$, convergent on the punctured plane $\mathbb C\setminus\{0\}$.
Whenever $\psi_n\neq0$ for infinitely many $n<0$, the generic case,
already realised by the fiducial state $\ket\Omega=\sum_j e^{-j^{2}/2}\ket j$, the point $\xi=0$ is, by the standard classification of isolated
singularities of a Laurent series, an essential singularity, not
a pole or a removable point. By the Weierstrass-Casorati theorem,
$G_\psi$ then takes values arbitrarily close to every complex number in
any punctured neighbourhood of $\xi=0$ (and symmetrically at
$\xi=\infty$). Hence $G_\psi(\xi)$ cannot be continued to an entire
function on $\mathbb C$, and the Weierstrass--Hadamard theorem, which
presupposes holomorphy on all of $\mathbb C$, cannot be invoked for it.
Only for finite-rank states, \textit{i.e}.\ $\psi_n$ supported on a finite range
of $n$, is $G_\psi(\xi)$ a genuine (Laurent) polynomial, entire
everywhere including at $\xi=0$.

\begin{figure*}
  \centering
    \includegraphics[width=\textwidth]{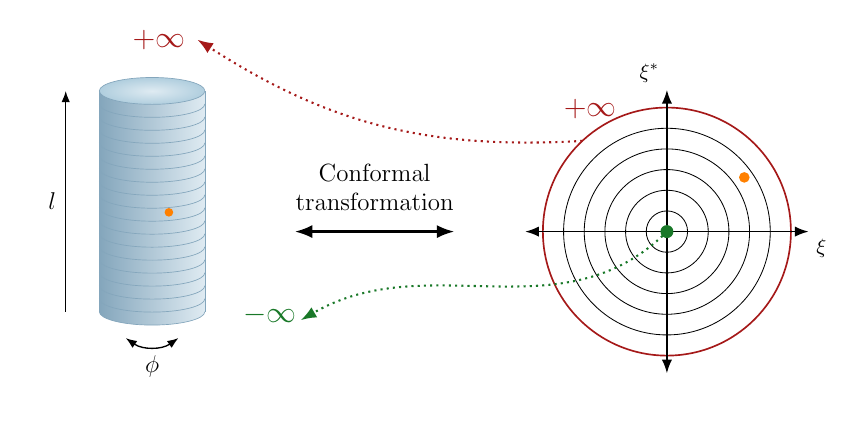}
  \caption{The exponential map $\xi=e^{z}=e^{l}e^{-i\phi}$ sends the discrete cylinder whose longitudinal coordinate $l$ and the angular coordinate $\phi$ onto the punctured complex plane $\mathbb{C}^{*}=\mathbb{C}\setminus\{0\}$.  Circles of constant $l$ are mapped to concentric circles of radius $e^{l}$ centered at the origin, while lines of constant $\phi$ are mapped to radial rays.  The negative infinite  $l\to-\infty$ and positive infinite  $l\to+\infty$ on the cylinder correspond respectively to the puncture at $\xi=0$ and the point at infinity, reflecting the infinite extent of the cylinder in the longitudinal direction.}
  \label{fig:conformal}
\end{figure*}

\subsubsection{Analytic properties of the stellar function}

For $|\psi_n|\le1$ and any fixed $z\in\mathbb C$, the terms of $G_\psi(z)=\sum_{n\in\mathbb Z}e^{-n^2/2}\psi_n e^{nz}$ obey $|e^{-n^2/2}\psi_n e^{nz}|\le e^{-n^2/2+n\,\mathrm{Re}(z)}$, which decays as $e^{-n^2/2}$ for $|n|\to\infty$ at fixed $\mathrm{Re}(z)$; by the Weierstrass $M$-test the series therefore converges absolutely and uniformly on compact subsets of $\mathbb C$. Consequently $G_\psi$ is entire on all of $\mathbb C$, $2\pi i$-periodic, and has no exceptional point (in particular none at $z=0$).

Its growth follows from completing the square,
\begin{equation}
\sum_{n\in\mathbb Z}e^{-n^{2}/2+nx}=e^{x^{2}/2}\sum_{n\in\mathbb Z}e^{-(n-x)^{2}/2}\sim\sqrt{2\pi}\,e^{x^{2}/2}\qquad(x\to\infty),
\end{equation}
so that $\log M(r)\sim r^{2}/2$ along the real axis. A sharper, uniformly valid bound is obtained from the Poisson summation formula,
\begin{equation}
\sum_{n\in\mathbb Z}e^{-(n-x)^{2}}=\sqrt\pi\sum_{k\in\mathbb Z}e^{-\pi^{2}k^{2}}e^{2\pi i kx},\qquad x\in\mathbb R,
\end{equation}
whose right-hand side, since $e^{-\pi^2}\approx3.5\times10^{-5}$, is extremely close to $\sqrt\pi$ for every $x$ and exactly maximal, equal to $\varpi=\vartheta_3(0,e^{-1})$, at integer $x$. Either estimate yields $\log\log M(r)/\log r\to2$: $G_\psi$ is of order $2$, mean type the growth class of a Jacobi theta function.\\

For the fiducial state $\ket\Omega=\sum_{n\in\mathbb{Z}} e^{-n^2/2}\ket n$,
\begin{equation}
G_\Omega(z)=\sum_{n\in\mathbb Z}e^{-n^{2}/2}e^{nz}=\vartheta_3\!\Big({-}\frac{iz}{2},\,e^{-1/2}\Big),
\end{equation}
using $\vartheta_3(w,q)=\sum_{n\in\mathbb{Z}} q^{n^2}e^{2inw}$ with $q=e^{-1/2}$, $w=z/(2i)$. With $z=\ell-i\phi$ this is precisely the theta function already appearing in the $\phi$-representation of the wrapped coherent state above: the two are the same analytic object. The classical zero lattice of $\vartheta_3$, $w=\tfrac\pi2+\tfrac\pi2\tau+m\pi+n\pi\tau$ with $q=e^{i\pi\tau}$ and $m,n\in\mathbb Z$, a single line for real arguments, thus governs the zero structure of $G_\Omega$.

\subsection{Holomorphic representation and stellar zeros}\label{app:counting}

The Bargmann or stellar representation maps this state onto a holomorphic function through its overlap with a coherent state $|\xi\rangle = \sum_{n\in\mathbb{Z}} c_n \xi^n |n\rangle$:
\begin{equation}\label{eq:bargmann_overlap}
H_\psi(\xi) = \langle \psi | \xi \rangle = \sum_{n \in \mathbb{Z}} \psi_n^* c_n \, \xi^n.
\end{equation}
For a finite superposition truncated at $n = N$, in the absence of singularity at $\xi=0$, $H_\psi(\xi)$ reduces to a polynomial of degree $N$. By the fundamental theorem of algebra it factorises over its distinct zeros $\xi_j$ with multiplicities $m_j$ as
\begin{equation}\label{eq:factorization}
H_\psi(\xi) = C \prod_{j=1}^{K} (\xi - \xi_j)^{m_j}, \qquad \sum_{j=1}^{K} m_j = N,
\end{equation}
where $C = \psi_N^* c_N$ is a normalization constant and $K$ is the number of distinct zeros. Denoting by $\xi_1,\dots,\xi_N$ the $N$ roots counted with multiplicity, expanding the product yields
\begin{equation}\label{eq:expanded_product}
\prod_{j=1}^{N}(\xi-\xi_j) = \sum_{k=0}^{N} (-1)^{N-k} e_{N-k}(\xi_1,\dots,\xi_N) \, \xi^k,
\end{equation}
with $e_k$ the elementary symmetric polynomials,
\begin{equation}\label{eq:symmetric_poly}
e_k(\xi_1,\dots,\xi_N) = \sum_{1 \leq j_1 < \cdots < j_k \leq N} \xi_{j_1} \cdots \xi_{j_k},
\end{equation}
and $e_0 = 1$. Matching coefficients between \eqref{eq:bargmann_overlap} and \eqref{eq:expanded_product} gives the general relation
\begin{equation}\label{eq:coefficient_matching}
\psi_k^* c_k = (-1)^{N-k} \, e_{N-k}(\xi_1,\dots,\xi_N), \qquad k = 0, 1, \dots, N.
\end{equation}
For example, for $N=2$ with simple zeros one finds $\psi_2^* c_2 = 1$, $\psi_1^* c_1 = -(\xi_1 + \xi_2)$, and $\psi_0^* c_0 = \xi_1 \xi_2$. Equation \eqref{eq:coefficient_matching} thus establishes a direct correspondence between the state coefficients and the stellar zeros.

Consider a state characterized by a single zero $\xi_0$, whose Bargmann function takes the form $(\xi - \xi_0)$. The angular-momentum coefficients can be extracted via contour integration. Choosing a circular contour $\gamma$ of radius $R$ and exploiting the radial dependence of the coherent-state normalization, one writes
\begin{equation}\label{eq:contour_single_zero}
\psi_n \, e^{-n^2/2} = \frac{1}{2\pi i} \oint_{\gamma} e^{(\ln|\xi|)^2/2} \, \frac{\xi - \xi_0}{\xi^{n+1}} \, d\xi.
\end{equation}
Since $|\xi| = R$ is constant on $\gamma$, the Gaussian factor $e^{(\ln|\xi|)^2/2} = e^{(\ln R)^2/2}$ can be pulled out of the integral. Using $\oint_\gamma \xi^m d\xi = 2\pi i \, \delta_{m,-1}$, Eq.~\eqref{eq:contour_single_zero} evaluates to
\begin{equation}\label{eq:single_zero_result}
\psi_n \, e^{-n^2/2} = e^{(\ln R)^2/2} \times
\begin{cases}
1, & n = 1, \\
- \xi_0, & n = 0, \\
0, & \text{otherwise}.
\end{cases}
\end{equation}
This confirms that a single zero in the Bargmann function yields a state supported on only two angular-momentum eigenstates, $n = 0$ and $n = 1$, with coefficients fixed by the zero location $\xi_0$.\\

\section{Operator algebra in the stellar representation}

\subsection{Cylindrical and plane coordinates}\label{sec:cylindrical}

In the cylindrical phase space $(l,\phi)$, the fundamental operators act on the stellar representation, or Bargmann states,  as follows \cite{Kowalski:1996mz}. The angular-momentum operator $\hat{L}$ generates translations in $\phi$,
\begin{equation}\label{eq:J_cyl}
\bra{\psi}\hat{L}\ket{l,\phi} = \frac{\partial}{\partial l}\psi(l,\phi),
\end{equation}
while the unitary shift $\hat{E}=e^{i\hat{\phi}}$ increments the angular momentum,
\begin{equation}\label{eq:U_cyl}
\bra{\psi}\hat{E}\ket{l,\phi} = e^{-(\mathrm{i}\phi-l)}e^{-1/2}\bra{\psi}\ket{l-1,\phi}.
\end{equation}
Its powers and the phase operator $\hat{X}=e^{\hat{\phi}-\mathrm{i}\hat{L}}$ satisfy
\begin{align}
\bra{\psi}\hat{E}^{m}\ket{l,\phi} &= e^{-m^{2}/2}e^{-(\mathrm{i}\phi-l)m}\bra{\psi}\ket{l-m,\phi}, \label{eq:Um_cyl}\\
\bra{\psi}\hat{X}\ket{l,\phi} &= e^{-(\mathrm{i}\phi-l)}\bra{\psi}\ket{l,\phi}. \label{eq:X_cyl}
\end{align}
The rotation operator $\hat{V}^{\phi_{0}}=e^{\mathrm{i}\phi_{0}\hat{L}}$ simply shifts the angle,
\begin{equation}\label{eq:V_cyl}
\bra{\psi}\hat{V}^{\phi_{0}}\ket{l,\phi} = \bra{\psi}\ket{l,\phi+\phi_{0}}.
\end{equation}
Combining these, the displacement operator $\hat{D}(m,\phi_{0})=\hat{E}^{m}e^{-\mathrm{i}\hat{L}\phi_{0}-\frac{1}{2}m\phi_{0}}$ (note the additional phase factor relative to $\hat{X}$) acts as
\begin{equation}\label{eq:displacement_cyl}
\bra{\psi}\hat{D}(m,\phi_{0})\ket{l,\phi} = e^{\frac{1}{2}m\phi_{0}} e^{-m(\mathrm{i}\phi-l)} e^{-m^{2}/2} \bra{\psi}\ket{l-m,\phi-\phi_{0}}.
\end{equation}
In the plane Bargmann representation, the operators take the form
\begin{equation}\label{eq:L_plane}
\hat{L}\,\phi(z^{*}) = -z^{*}\frac{d}{dz^{*}}\phi(z^{*}),
\end{equation}
and the shift operators become
\begin{equation}\label{eq:U_plane}
\hat{E}\,\phi(z^{*}) = \frac{\phi(ez^{*})}{\sqrt{e}\,z^{*}}, \qquad
\hat{E}^{\dagger}\phi(z^{*}) = \frac{z^{*}}{\sqrt{e}}\phi(e^{-1}z^{*}).
\end{equation}
Their matrix elements on coherent states are
\begin{align}
\bra{\psi}\hat{E}^{m}\ket{z} &= e^{-m^{2}/2}z^{-m}\bra{\psi}\ket{ze^{m}}, \label{eq:Um_plane}\\
\bra{\psi}\hat{V}^{\phi_{0}}\ket{z} &= \bra{\psi}\ket{ze^{\mathrm{i}\phi_{0}}}, \label{eq:V_plane}
\end{align}
so that the full displacement operator reads
\begin{equation}\label{eq:displacement_plane}
\bra{\psi}\hat{D}(\phi_{0},m)\ket{z} = e^{-\frac{1}{2}m\phi_{0}}e^{-m^{2}/2}(ze^{\mathrm{i}\phi_{0}})^{-m}\bra{\psi}\ket{ze^{\mathrm{i}\phi_{0}}e^{m}}.
\end{equation}

Equation \eqref{eq:displacement_plane} reveals that a displacement with $m\neq0$ creates a zero of multiplicity $|m|$ at the origin (for $m<0$) or at infinity (for $m>0$), while shifting the remaining zeros. In the cylindrical representation \eqref{eq:displacement_cyl}, however, these zeros at $z=0$ or $z\to\infty$ are not visible; they correspond to the two ``infinities'' of the plane that are mapped to the boundary of the cylinder. Hence, displacements in the plane can appear to change the zero count when viewed on the cylinder, but this is merely an artifact of the coordinate mapping.

\subsection{Free evolution and conservation of zeros}\label{sec:conservation}
The free-particle Hamiltonian on the circle, $\hat{H}=\nu\hat{L}^{2}$, generates the evolution
\begin{equation}\label{eq:free_evolution}
\bra{\psi}e^{-\mathrm{i}\nu\hat{L}^{2}}\ket{\xi} = \sum_{n\in\mathbb{Z}}\psi_n^*\,c_{n}\,e^{-\mathrm{i}\nu n^{2}}\,\xi^{n}.
\end{equation}
The zeros of the third Jacobi theta function are given by $\xi_{nm}=\exp\!\bigl(n\tau+m+\tfrac{\tau+1}{2}\bigr)$, with here $\mathrm{i}\pi\tau=-\frac{1}{2}-\mathrm{i}\nu$. Multiplying the coefficients by a unimodular phase $e^{-\mathrm{i}\nu n^{2}}$ does not change the degree of the polynomial, and hence the total number of zeros (counted with multiplicities) remains the same. However, the locations of the zeros in the complex $\xi$-plane may shift because the relative phases among the coefficients are modified.

If the original overlap function is
\begin{equation}\label{eq:F_psi_original}
H_{\psi}(\xi)=\sum_{n=0}^{N}\psi_{n}^{*}c_{n}\,\xi^{n} = \prod_{j=1}^{K}(\xi-\xi_{j})^{m_{j}}, \qquad \sum_{j=1}^{K}m_{j}=N,
\end{equation}
then under the free evolution the coefficients transform as $c_{n}\to c_{n}e^{-\mathrm{i}\nu n^{2}}$, and the modified function becomes
\begin{equation}\label{eq:F_psi_tilde}
\widetilde{H}_{\psi}(\xi)=\sum_{n=0}^{N}\psi_{n}^{*}\,c_{n}e^{-\mathrm{i}\nu n^{2}}\,\xi^{n}.
\end{equation}
This new function is still a polynomial of the same degree $N$ (assuming the same number of nonzero terms), so by the fundamental theorem of algebra it has the same number of zeros counted with multiplicities. The phase factors $e^{-\mathrm{i}\nu n^{2}}$ can change the relative interference among the terms, thereby changing the precise locations of the zeros, but not their count.

Denoting by $\xi_{1},\dots,\xi_{N}$ the $N$ zeros of $H_{\psi}$ counted with multiplicity, the general identification for the coefficients reads
\begin{equation}\label{eq:coeff_id}
\psi_{k}^{*}c_{k}=(-1)^{N-k}\,e_{N-k}(\xi_{1},\dots,\xi_{N}),\qquad k=0,1,\dots,N.
\end{equation}
Inserting the phase factors gives
\begin{equation}\label{eq:coeff_id_phased}
\psi_{k}^{*}\,c_{k}e^{-\mathrm{i}\nu k^{2}}=(-1)^{N-k}\,e_{N-k}(\tilde{\xi}_{1},\dots,\tilde{\xi}_{N}),
\end{equation}
where $\tilde{\xi}_{1},\dots,\tilde{\xi}_{N}$ are the zeros of the evolved function $\widetilde{H}_{\psi}$. Thus, while the phases in the coefficients are modified, the number of roots (which is $N$) does not change.

More generally, a transformation $e^{-\mathrm{i}\nu\hat{L}^{p}}$ does not modify the number of zeros. Multiplying each coefficient by a phase factor
\begin{equation}\label{eq:phase_factor}
c_{n}\to c_{n}\,e^{-\mathrm{i}\nu n^{p}}
\end{equation}
(with any exponent $p$) is a unimodular change (the factor has absolute value 1). The polynomial
\begin{equation}\label{eq:F_psi_general}
H_{\psi}(\xi)=\sum_{n=0}^{N}\psi_{n}^{*}\,c_{n}\,\xi^{n}
\end{equation}
then gets modified to
\begin{equation}\label{eq:F_psi_modified}
\widetilde{H}_{\psi}(\xi)=\sum_{n=0}^{N}\psi_{n}^{*}\,c_{n}\,e^{-\mathrm{i}\nu n^{p}}\,\xi^{n}.
\end{equation}
Since each term still appears with the same power $\xi^{n}$ and the overall degree remains $N$ (provided none of the modified coefficients vanish by accident), the total number of zeros counted with multiplicity does not change. However, the locations of the zeros in the complex $\xi$-plane will in general shift because the relative phases of the coefficients have been altered.

\subsection{Transformations that change the number of zeros}\label{app:modzero}

Consider the operator $\hat{X}$ defined by
\begin{equation}\label{eq:X_plane}
\bra{\xi}\hat{X}\ket{\phi} = \frac{\phi(e^{2}\xi^{*})}{e\,\xi^{*}}, \qquad
\bra{\xi}\hat{X}^{\dagger}\ket{\phi} = \xi^{*}\phi(\xi^{*}).
\end{equation}
In the plane, $\hat{X}^{\dagger}$ introduces a zero at the origin (or infinity under $\hat{X}$), but when mapped to cylinder coordinates these points lie at the boundary and do not correspond to genuine zeros of the stellar function. Thus $\hat{X}$ does not, by itself, change the number of physical zeros.

A transformation that does modify the zero count is one that changes the polynomial degree. This occurs, for instance, when high- or low-order coefficients are set to zero by truncation or projection, which reduces the effective degree $N$ and thereby the number of zeros. Similarly, multiplying coefficients by $n$-dependent factors that are not pure phases which are a non-unimodular scaling can annihilate certain terms and alter the degree. A more dynamical example is provided by a unitary of the form $\exp[\mathrm{i}\lambda\cos(\hat{\phi})]$, which couples angular-momentum eigenstates. Its action on $\ket{l}$ reads \cite{PhysRevResearch.5.L032006}
\begin{equation}\label{eq:cos_hamiltonian}
e^{\mathrm{i}\lambda\cos(\hat{\phi})t}\ket{l} = \sum_{l'\in\mathbb{Z}} \mathrm{i}^{l'}J_{l'}(\lambda t)\ket{l+l'},
\end{equation}
where $J_{l'}$ are Bessel functions. Because the sum extends over an infinite range of $l'$, the polynomial degree changes and the number of zeros is no longer conserved. \\

Finally, a M\"obius transformation of $\xi \mapsto (a\xi+b)/(c\xi+d)$ generates a rank changing transformation, and we leave it for future work.

\bibliography{Cylinderstellar}

@article{Majorana:1932ul,
	author = {E. Majorana},
	da = {1932//},
	doi = {10.1007/BF02960953},
	id = {Majorana1932},
	isbn = {1827-6121},
	journal = {Nuovo Cimento},
	number = {2},
	pages = {43--50},
	title = {Atomi orientati in campo magnetico variabile},
	ty = {JOUR},
	url = {http://dx.doi.org/10.1007/BF02960953},
	volume = {9},
	year = {1932}}

@article{rehacek_experimental_2008,
	title = {Experimental test of uncertainty relations for quantum mechanics on a circle},
	volume = {77},
	issn = {1050-2947, 1094-1622},
	url = {http://arxiv.org/abs/0712.0230},
	doi = {10.1103/PhysRevA.77.032110},
	pages = {032110},
	number = {3},
	journaltitle = {Physical Review A},
	journal = {Phys. Rev. A},
	author = {Rehacek, J. and Bouchal, Z. and Celechovsky, R. and Hradil, Z. and Sanchez-Soto, L. L.},
	urldate = {2026-08-05},
	year = {2008},
	langid = {english},
	eprinttype = {arxiv},
	eprint = {0712.0230 [quant-ph]},
}

@article{franke-arnold_uncertainty_2004,
	title = {Uncertainty principle for angular position and angular momentum},
	volume = {6},
	issn = {1367-2630},
	url = {https://iopscience.iop.org/article/10.1088/1367-2630/6/1/103},
	doi = {10.1088/1367-2630/6/1/103},
	pages = {103--103},
	journaltitle = {New Journal of Physics},
	journal = {New J. Phys.},
	author = {Franke-Arnold, Sonja and Barnett, Stephen M and Yao, Eric and Leach, Jonathan and Courtial, Johannes and Padgett, Miles},
	urldate = {2026-08-05},
	year = {2004},
	langid = {english},
}

@article{pegg_minimum_2005,
	title = {Minimum uncertainty states of angular momentum and angular position},
	volume = {7},
	issn = {1367-2630},
	url = {https://iopscience.iop.org/article/10.1088/1367-2630/7/1/062},
	doi = {10.1088/1367-2630/7/1/062},
	pages = {62--62},
	journaltitle = {New Journal of Physics},
	journal = {New J. Phys.},
	author = {Pegg, David T and Barnett, Stephen M and Zambrini, Roberta and Franke-Arnold, Sonja and Padgett, Miles},
	urldate = {2026-08-05},
	year = {2005},
	langid = {english},
}

@book{Bacry:2004aa,
	address = {Paris},
	author = {H. Bacry},
	publisher = {Publibook},
	title = {Group {T}heory and {C}onstellations},
	year = {2004}}

@article{forbes_quantum_2019,
	title = {Quantum mechanics with patterns of light: {Progress} in high dimensional and multidimensional entanglement with structured light},
	volume = {1},
	issn = {2639-0213},
	shorttitle = {Quantum mechanics with patterns of light},
	url = {https://pubs.aip.org/aqs/article/1/1/011701/997233/Quantum-mechanics-with-patterns-of-light-Progress},
	doi = {10.1116/1.5112027},
	number = {1},
	urldate = {2026-07-28},
	journal = {AVS Quantum Science},
	author = {Forbes, Andrew and Nape, Isaac},
	month = dec,
	year = {2019},
	pages = {011701},
}

@misc{Ginsparg:1988ui,
	title = {Applied {Conformal} {Field} {Theory}},
	url = {http://arxiv.org/abs/hep-th/9108028},
	doi = {10.48550/arXiv.hep-th/9108028},
	urldate = {2026-07-29},
	publisher = {arXiv},
	author = {Ginsparg, Paul},
	month = nov,
	year = {1988},
	note = {arXiv:hep-th/9108028},
}

@book{DiFrancesco:1997nk,
  title     = {Conformal Field Theory},
  author    = {Francesco, Philippe and Mathieu, Pierre and S{\'e}n{\'e}chal, David},
  year      = {1997},
  publisher = {Springer},
  address   = {New York, NY},
  series    = {Graduate Texts in Contemporary Physics},
  isbn      = {978-0-387-94785-3},
  doi       = {10.1007/978-1-4612-2256-9},
  pages     = {890},
  edition   = {1},
  issn      = {0938-037X}
}

@article{Belavin:1984vu,
title = {Infinite conformal symmetry in two-dimensional quantum field theory},
journal = {Nuclear Physics B},
volume = {241},
number = {2},
pages = {333-380},
year = {1984},
issn = {0550-3213},
doi = {https://doi.org/10.1016/0550-3213(84)90052-X},
url = {https://www.sciencedirect.com/science/article/pii/055032138490052X},
author = {A.A. Belavin and A.M. Polyakov and A.B. Zamolodchikov}
}

@article{fan_cooper-pair_2006,
	title = {Cooper-pair number–phase {Wigner} function for the bosonic operator {Josephson} model},
	volume = {359},
	copyright = {https://www.elsevier.com/tdm/userlicense/1.0/},
	issn = {03759601},
	url = {https://linkinghub.elsevier.com/retrieve/pii/S0375960106010875},
	doi = {10.1016/j.physleta.2006.07.007},
	number = {6},
	urldate = {2026-07-28},
	journal = {Physics Letters A},
	author = {Fan, Hong-Yi and Wang, Ji-Suo and Liu, Shu-guang},
	month = dec,
	year = {2006},
	pages = {580--586},
}

@article{dirienzo_coupled_1983,
	title = {A coupled angular momentum model for the {Josephson} junction},
	volume = {51},
	issn = {0002-9505, 1943-2909},
	url = {https://pubs.aip.org/ajp/article/51/7/587/1052182/A-coupled-angular-momentum-model-for-the-Josephson},
	doi = {10.1119/1.13172},
	number = {7},
	urldate = {2026-07-28},
	journal = {American Journal of Physics},
	author = {DiRienzo, Andrew L. and Young, Richard A.},
	month = jul,
	year = {1983},
	pages = {587--597},
}

@article{ma_realization_2020,
	title = {Realization of all-optical vortex switching in exciton-polariton condensates},
	volume = {11},
	issn = {2041-1723},
	url = {https://www.nature.com/articles/s41467-020-14702-5},
	doi = {10.1038/s41467-020-14702-5},
	number = {1},
	urldate = {2026-07-28},
	journal = {Nature Communications},
	author = {Ma, Xuekai and Berger, Bernd and Aßmann, Marc and Driben, Rodislav and Meier, Torsten and Schneider, Christian and Höfling, Sven and Schumacher, Stefan},
	month = feb,
	year = {2020},
	pages = {897},
}

@article{PhysRevA.95.052111,
  title = {Wigner functions for angle and orbital angular momentum: Operators and dynamics},
  author = {Kastrup, H. A.},
  journal = {Phys. Rev. A},
  volume = {95},
  issue = {5},
  pages = {052111},
  numpages = {13},
  year = {2017},
  month = {May},
  publisher = {American Physical Society},
  doi = {10.1103/PhysRevA.95.052111},
  url = {https://link.aps.org/doi/10.1103/PhysRevA.95.052111}
}

@article{bruno_quantum_2012,
	title = {Quantum geometric phase in {Majorana}'s stellar representation: {Mapping} onto a many-body {Aharonov}-{Bohm} phase},
	volume = {108},
	issn = {0031-9007, 1079-7114},
	shorttitle = {Quantum geometric phase in {Majorana}'s stellar representation},
	url = {http://arxiv.org/abs/1204.2372},
	doi = {10.1103/PhysRevLett.108.240402},
	number = {24},
	urldate = {2022-12-18},
	journal = {Physical Review Letters},
	author = {Bruno, Patrick},
	month = jun,
	year = {2012},
	note = {arXiv:1204.2372 [cond-mat, physics:math-ph, physics:quant-ph]},
	pages = {240402},
}

@article{floerchinger_wehrl_2021,
	title = {Wehrl entropy, entropic uncertainty relations and entanglement},
	volume = {103},
	issn = {2469-9926, 2469-9934},
	url = {http://arxiv.org/abs/2103.07229},
	doi = {10.1103/PhysRevA.103.062222},
	number = {6},
	urldate = {2026-07-28},
	journal = {Physical Review A},
	author = {Floerchinger, Stefan and Haas, Tobias and Müller-Groeling, Henrik},
	month = jun,
	year = {2021},
	note = {arXiv:2103.07229 [quant-ph]},
	pages = {062222},
}

@article{RevModPhys.50.221,
  title = {General properties of entropy},
  author = {Wehrl, Alfred},
  journal = {Rev. Mod. Phys.},
  volume = {50},
  issue = {2},
  pages = {221--260},
  numpages = {0},
  year = {1978},
  month = {Apr},
  publisher = {American Physical Society},
  doi = {10.1103/RevModPhys.50.221},
  url = {https://link.aps.org/doi/10.1103/RevModPhys.50.221}
}

@article{nieto_wigner_1998,
	title = {Wigner distribution function for {Euclidean} systems},
	volume = {31},
	issn = {0305-4470, 1361-6447},
	url = {https://iopscience.iop.org/article/10.1088/0305-4470/31/16/015},
	doi = {10.1088/0305-4470/31/16/015},
	number = {16},
	urldate = {2022-12-21},
	journal = {Journal of Physics A: Mathematical and General},
	author = {Nieto, Luis Miguel and Atakishiyev, Natig M and Chumakov, Sergey M and Wolf, Kurt Bernardo},
	month = apr,
	year = {1998},
	pages = {3875--3895},
}

@article{sedov_circular_2021,
	title = {Circular polariton currents with integer and fractional orbital angular momenta},
	volume = {3},
	issn = {2643-1564},
	url = {https://link.aps.org/doi/10.1103/PhysRevResearch.3.013072},
	doi = {10.1103/PhysRevResearch.3.013072},
	number = {1},
	urldate = {2026-07-28},
	journal = {Physical Review Research},
	author = {Sedov, E. S. and Lukoshkin, V. A. and Kalevich, V. K. and Savvidis, P. G. and Kavokin, A. V.},
	month = jan,
	year = {2021},
	pages = {013072},
}

@article{Yao:11,
author = {Alison M. Yao and Miles J. Padgett},
journal = {Adv. Opt. Photon.},
number = {2},
pages = {161--204},
publisher = {Optica Publishing Group},
title = {Orbital angular momentum: origins, behavior and applications},
volume = {3},
month = {Jun},
year = {2011},
url = {https://opg.optica.org/aop/abstract.cfm?URI=aop-3-2-161},
doi = {10.1364/AOP.3.000161},
}

@article{molina-terriza_twisted_2007,
	title = {Twisted photons},
	volume = {3},
	issn = {1745-2481},
	url = {https://doi.org/10.1038/nphys607},
	doi = {10.1038/nphys607},
	number = {5},
	journal = {Nature Physics},
	author = {Molina-Terriza, Gabriel and Torres, Juan P. and Torner, Lluis},
	month = may,
	year = {2007},
	pages = {305--310},
}

@article{PhysRevLett.97.170406,
  title = {Quantized Rotation of Atoms from Photons with Orbital Angular Momentum},
  author = {Andersen, M. F. and Ryu, C. and Clad\'e, Pierre and Natarajan, Vasant and Vaziri, A. and Helmerson, K. and Phillips, W. D.},
  journal = {Phys. Rev. Lett.},
  volume = {97},
  issue = {17},
  pages = {170406},
  numpages = {4},
  year = {2006},
  month = {Oct},
  publisher = {American Physical Society},
  doi = {10.1103/PhysRevLett.97.170406},
  url = {https://link.aps.org/doi/10.1103/PhysRevLett.97.170406}
}

@article{lakshminarayan_entanglement_2016,
	title = {Entanglement and localization transitions in eigenstates of interacting chaotic systems},
	volume = {94},
	issn = {2470-0045, 2470-0053},
	url = {http://arxiv.org/abs/1601.07061},
	doi = {10.1103/PhysRevE.94.010205},
	pages = {010205},
	number = {1},
	journaltitle = {Physical Review E},
	journal = {Phys. Rev. E},
	author = {Lakshminarayan, Arul and Srivastava, Shashi C. L. and Ketzmerick, Roland and Bäcker, Arnd and Tomsovic, Steven},
	urldate = {2025-03-06},
	year = {2016},
	langid = {english},
	eprinttype = {arxiv},
	eprint = {1601.07061 [quant-ph]},
}

@article{flambaum_entropy_2001,
	title = {Entropy production and wave packet dynamics in the Fock space of closed chaotic many-body systems},
	volume = {64},
	issn = {1063-651X, 1095-3787},
	url = {http://arxiv.org/abs/quant-ph/0103129},
	doi = {10.1103/PhysRevE.64.036220},
	pages = {036220},
	number = {3},
	journal = {Physical Review E},
	shortjournal = {Phys. Rev. E},
	author = {Flambaum, V. V. and Izrailev, F. M.},
	urldate = {2025-03-06},
	year= {2001},
	langid = {english},
	eprinttype = {arxiv},
	eprint = {quant-ph/0103129},
}

@article{calixto_inverse_2015,
	title = {Inverse participation ratio and localization in topological insulator phase transitions},
	volume = {2015},
	issn = {1742-5468},
	url = {http://arxiv.org/abs/1610.03696},
	doi = {10.1088/1742-5468/2015/06/P06029},
	pages = {P06029},
	number = {6},
	journaltitle = {Journal of Statistical Mechanics: Theory and Experiment},
	journal = {J. Stat. Mech.},
	author = {Calixto, M. and Romera, E.},
	urldate = {2025-03-06},
	year= {2015},
	langid = {english},
	eprinttype = {arxiv},
	eprint = {1610.03696 [cond-mat]},
}

@article{misguich_inverse_2016,
	title = {Inverse participation ratios in the {XXZ} spin chain},
	volume = {94},
	rights = {http://link.aps.org/licenses/aps-default-license},
	issn = {2469-9950, 2469-9969},
	url = {https://link.aps.org/doi/10.1103/PhysRevB.94.155110},
	doi = {10.1103/PhysRevB.94.155110},
	pages = {155110},
	number = {15},
	journaltitle = {Physical Review B},
	journal = {Phys. Rev. B},
	author = {Misguich, Grégoire and Pasquier, Vincent and Luck, Jean-Marc},
	urldate = {2025-03-06},
	year= {2016},
	langid = {english},
}

@article{PhysRevA.74.053809,
  title = {Entanglement of orbital angular momentum states between an ensemble of cold atoms and a photon},
  author = {Inoue, R. and Kanai, N. and Yonehara, T. and Miyamoto, Y. and Koashi, M. and Kozuma, M.},
  journal = {Phys. Rev. A},
  volume = {74},
  issue = {5},
  pages = {053809},
  numpages = {5},
  year = {2006},
  month = {Nov},
  publisher = {American Physical Society},
  doi = {10.1103/PhysRevA.74.053809},
  url = {https://link.aps.org/doi/10.1103/PhysRevA.74.053809}
}

@article{Gazeau:2022aa,
	author = {Gazeau, J.-P. and Murenzi, R.},
	doi = {https://doi.org/10.3390/quantum4040026},
	journal = {Quantum Rep.},
	pages = {362-379},
	title = {Integral Quantization for the Discrete Cylinder},
	volume = {4},
	year = {2022}}

@article{Dirichlet:1829aa,
	author = {Dirichlet, P.G.L},
	doi = {https://doi.org/10.1515/crll.1829.4.157},
	journal = {J. f{\"u}r Math.},
	pages = {157--169},
	title = {Sur la convergence des s{\'e}ries trigonometriques qui servent {\`a} repr{\'e}senter une fonction arbitraire entre des limites donn{\'e}es},
	volume = {4},
	year = {1829}}

@article{Fejer:1904aa,
	author = {Fej{\'e}r, Leopold},
	doi = {http://eudml.org/doc/158116},
	journal = {Math. Ann.},
	number = {51--69},
	title = {Untersuchungen {\"u}ber {F}ouriersche {R}eihen},
	volume = {58},
	year = {1904}}

@book{Humphreys:1972aa,
	address = {NewYork},
	author = {Humphreys, J. E.},
	publisher = {Springer},
	title = {Introduction to {L}ie Algebras and Representation Theory},
	year = {1972}}

@article{Lerner:1968aa,
	author = {Lerner, E. C.},
	date = {1968/07/01},
	doi = {10.1007/BF02711966},
	id = {Lerner1968},
	isbn = {1826-9877},
	journal = {Nuovo Cimento B},
	number = {1},
	pages = {183--186},
	title = {Harmonic-oscillator phase operators},
	url = {https://doi.org/10.1007/BF02711966},
	volume = {56},
	year = {1968}}

@article{Hradil:2006aa,
	author = {Hradil, Z. and {\v R}eh{\'a}{\v c}ek, J. and Bouchal, Z. and {\v C}elechovsk{\'y}, R. and S{\'a}nchez-Soto, L. L.},
	date = {2006/12/15/},
	day = {15},
	doi = {10.1103/PhysRevLett.97.243601},
	id = {10.1103/PhysRevLett.97.243601},
	j1 = {PRL},
	journal = {Phys. Rev. Lett.},
	journal1 = {Phys. Rev. Lett.},
	month = {12},
	number = {24},
	pages = {243601},
	publisher = {American Physical Society},
	title = {Minimum Uncertainty Measurements of Angle and Angular Momentum},
	url = {https://link.aps.org/doi/10.1103/PhysRevLett.97.243601},
	volume = {97},
	year = {2006}}

@article{Prosen1995QSOS,
  author       = {Toma{\v z} Prosen},
  title        = {Quantum Surface of Section Method: Eigenstates and Unitary Quantum Poincar{\'e} Evolution},
  journal      = {arXiv preprint chao-dyn/9503008},
  year         = {1995},
  eprint       = {chao-dyn/9503008},
  archivePrefix= {arXiv},
  primaryClass = {chao-dyn},
  note         = {CAMTP Preprint CAMTP/95-2},
}

@misc{nonnenmacher_chaotic_2006,
	title = {Chaotic eigenfunctions in phase space},
	url = {http://arxiv.org/abs/chao-dyn/9711016},
	doi = {10.48550/arXiv.chao-dyn/9711016},
	number = {{arXiv}:chao-dyn/9711016},
	publisher = {{arXiv}},
	author = {Nonnenmacher, S. and Voros, A.},
	urldate = {2026-07-25},
	year = {2006},
	langid = {english},
	eprinttype = {arxiv},
	eprint = {chao-dyn/9711016},
}

@article{leboeuf_chaos-revealing_1990,
	title = {Chaos-revealing multiplicative representation of quantum eigenstates},
	volume = {23},
	issn = {0305-4470, 1361-6447},
	url = {https://iopscience.iop.org/article/10.1088/0305-4470/23/10/017},
	doi = {10.1088/0305-4470/23/10/017},
	pages = {1765--1774},
	number = {10},
	journaltitle = {Journal of Physics A: Mathematical and General},
	journal = {J. Phys. A: Math. Gen.},
	author = {Leboeuf, P and Voros, A},
	urldate = {2024-11-22},
	year = {1990},
}

@article{Mista:2022tv,
	author = {Mi{\v s}ta, Ladislav and de Guise, Hubert and {\v R}eh{\'a}{\v c}ek, Jaroslav and Hradil, Zden{\v e}k},
	date = {2022/08/05/},
	day = {05},
	doi = {10.1103/PhysRevA.106.022204},
	id = {10.1103/PhysRevA.106.022204},
	j1 = {PRA},
	journal = {Phys. Rev. A},
	journal1 = {Phys. Rev. A},
	month = {08},
	number = {2},
	pages = {022204},
	publisher = {American Physical Society},
	title = {Angle and angular momentum: Uncertainty relations, simultaneous measurement, and phase-space representation},
	url = {https://link.aps.org/doi/10.1103/PhysRevA.106.022204},
	volume = {106},
	year = {2022}}

@article{PhysRevResearch.5.L032006,
  title = {Majorana stellar representation of twisted photons},
  author = {Fabre, Nicolas and Klimov, Andrei B. and Murenzi, Romain and Gazeau, Jean-Pierre and S\'anchez-Soto, Luis L.},
  journal = {Phys. Rev. Res.},
  volume = {5},
  issue = {3},
  pages = {L032006},
  numpages = {7},
  year = {2023},
  month = {Jul},
  publisher = {American Physical Society},
  doi = {10.1103/PhysRevResearch.5.L032006},
  url = {https://link.aps.org/doi/10.1103/PhysRevResearch.5.L032006}
}

@article{PhysRevA.97.032346,
  title = {Performance and structure of single-mode bosonic codes},
  author = {Albert, Victor V. and Noh, Kyungjoo and Duivenvoorden, Kasper and Young, Dylan J. and Brierley, R. T. and Reinhold, Philip and Vuillot, Christophe and Li, Linshu and Shen, Chao and Girvin, S. M. and Terhal, Barbara M. and Jiang, Liang},
  journal = {Phys. Rev. A},
  volume = {97},
  issue = {3},
  pages = {032346},
  numpages = {30},
  year = {2018},
  month = {Mar},
  publisher = {American Physical Society},
  doi = {10.1103/PhysRevA.97.032346},
  url = {https://link.aps.org/doi/10.1103/PhysRevA.97.032346}
}

@article{PhysRevLett.109.230503,
  title = {Positive Wigner Functions Render Classical Simulation of Quantum Computation Efficient},
  author = {Mari, A. and Eisert, J.},
  journal = {Phys. Rev. Lett.},
  volume = {109},
  issue = {23},
  pages = {230503},
  numpages = {5},
  year = {2012},
  month = {Dec},
  publisher = {American Physical Society},
  doi = {10.1103/PhysRevLett.109.230503},
  url = {https://link.aps.org/doi/10.1103/PhysRevLett.109.230503}
}

@article{Baragiola:2019,
  author  = {Baragiola, B. Q. and Pantaleoni, G. and Alexander, R. N. and Karanjai, A. and Menicucci, N. C.},
  title   = {All-{G}aussian Universality and Fault Tolerance with the {G}ottesman-{K}itaev-{P}reskill Code},
  journal = {Phys. Rev. Lett.},
  volume  = {123},
  pages   = {200502},
  year    = {2019},
  doi     = {10.1103/PhysRevLett.123.200502}
}

@article{Yamasaki:2020,
  author  = {Yamasaki, H. and Matsuura, T. and Koashi, M.},
  title   = {Cost-reduced all-{G}aussian universality with the {G}ottesman-{K}itaev-{P}reskill code: {R}esource-theoretic approach to cost analysis},
  journal = {Phys. Rev. Research},
  volume  = {2},
  pages   = {023270},
  year    = {2020},
  doi     = {10.1103/PhysRevResearch.2.023270}
}

@article{Bourassa:2021,
  author  = {Bourassa, J. E. and Alexander, R. N. and Vasmer, M. and Patil, A. and Tzitrin, I. and Matsuura, T. and Su, D. and Baragiola, B. Q. and Guha, S. and Dauphinais, G. and Sabapathy, K. K. and Menicucci, N. C. and Dhand, I.},
  title   = {Blueprint for a Scalable Photonic Fault-Tolerant Quantum Computer},
  journal = {Quantum},
  volume  = {5},
  pages   = {392},
  year    = {2021},
  doi     = {10.22331/q-2021-02-04-392}
}

@article{Noh:2022,
  author  = {Noh, K. and Chamberland, C. and Brand{\~a}o, F. G. S. L.},
  title   = {Low-overhead fault-tolerant quantum error correction with the surface-{GKP} code},
  journal = {PRX Quantum},
  volume  = {3},
  pages   = {010315},
  year    = {2022},
  doi     = {10.1103/PRXQuantum.3.010315}
}

@article{Royer:2020,
  author  = {Royer, B. and Singh, S. and Girvin, S. M.},
  title   = {Stabilization of finite-energy {G}ottesman-{K}itaev-{P}reskill states},
  journal = {Phys. Rev. Lett.},
  volume  = {125},
  pages   = {260509},
  year    = {2020},
  doi     = {10.1103/PhysRevLett.125.260509}
}

@article{Mirrahimi:2014,
  author  = {Mirrahimi, M. and Leghtas, Z. and Albert, V. V. and Touzard, S. and Schoelkopf, R. J. and Jiang, L. and Devoret, M. H.},
  title   = {Dynamically protected cat-qubits: a new paradigm for universal quantum computation},
  journal = {New J. Phys.},
  volume  = {16},
  pages   = {045014},
  year    = {2014},
  doi     = {10.1088/1367-2630/16/4/045014}
}

@article{Rosenblum:2018,
  author  = {Rosenblum, S. and Reinhold, P. and Mirrahimi, M. and Jiang, L. and Frunzio, L. and Schoelkopf, R. J.},
  title   = {Fault-tolerant detection of a quantum error},
  journal = {Science},
  volume  = {361},
  pages   = {266--270},
  year    = {2018},
  doi     = {10.1126/science.aat3996}
}

@article{Schlegel:2022,
  author  = {Schlegel, D. S. and Minganti, F. and Savona, V.},
  title   = {Quantum error correction using squeezed {S}chr\"odinger cat states},
  journal = {Phys. Rev. A},
  volume  = {106},
  pages   = {022431},
  year    = {2022},
  doi     = {10.1103/PhysRevA.106.022431}
}

@article{Albarelli:2018,
  author  = {Albarelli, F. and Genoni, M. G. and Paris, M. G. A. and Ferraro, A.},
  title   = {Resource theory of quantum non-{G}aussianity and {W}igner negativity},
  journal = {Phys. Rev. A},
  volume  = {98},
  pages   = {052350},
  year    = {2018},
  doi     = {10.1103/PhysRevA.98.052350}
}

@article{Kowalski:1996mz,
	author = {K. Kowalski and J. Rembieli{\'n}ski and L. C. Papaloucas},
	date = {1996/07/21},
	doi = {10.1088/0305-4470/29/14/034},
	isbn = {0305-4470; 1361-6447},
	journal = {J. Phys. A: Math. Gen.},
	number = {14},
	pages = {4149-4167},
	publisher = {IOP Publishing},
	title = {Coherent states for a quantum particle on a circle},
	url = {http://dx.doi.org/10.1088/0305-4470/29/14/034},
	volume = {29},
	year = {1996}}

@article{Ruzzi:2006dq,
	author = {M. Ruzzi and M. A. Marchiolli and E. C. da Silva and D. Galetti},
	date = {2006/07/19},
	doi = {10.1088/0305-4470/39/31/016},
	isbn = {0305-4470; 1361-6447},
	journal = {J. Phys. A: Math. Gen.},
	number = {31},
	pages = {9881--9890},
	publisher = {IOP Publishing},
	title = {{Quasiprobability distribution functions for periodic phase spaces: I. Theoretical aspects}},
	url = {http://dx.doi.org/10.1088/0305-4470/39/31/016},
	volume = {39},
	year = {2006}}

@article{gottesman_encoding_2001,
	title = {Encoding a qubit in an oscillator},
	volume = {64},
	issn = {1050-2947, 1094-1622},
	url = {http://arxiv.org/abs/quant-ph/0008040},
	doi = {10.1103/PhysRevA.64.012310},
	pages = {012310},
	number = {1},
	journaltitle = {Physical Review A},
	journal = {Phys. Rev. A},
	author = {Gottesman, Daniel and Kitaev, Alexei and Preskill, John},
	urldate = {2024-08-27},
	year = {2001},
	langid = {english},
	eprinttype = {arxiv},
	eprint = {quant-ph/0008040},
}

@article{Wigner:1932,
  author  = {Wigner, E. P.},
  title   = {On the Quantum Correction For Thermodynamic Equilibrium},
  journal = {Phys. Rev.},
  volume  = {40},
  pages   = {749--759},
  year    = {1932},
  doi     = {10.1103/PhysRev.40.749}
}

@article{Cahill:1969,
  author  = {Cahill, K. E. and Glauber, R. J.},
  title   = {Density Operators and Quasiprobability Distributions},
  journal = {Phys. Rev.},
  volume  = {177},
  pages   = {1882--1902},
  year    = {1969},
  doi     = {10.1103/PhysRev.177.1882}
}

@article{Kastrup:2016,
  author  = {Kastrup, H. A.},
  title   = {Wigner functions for the pair angle and orbital angular momentum},
  journal = {Phys. Rev. A},
  volume  = {94},
  pages   = {062113},
  year    = {2016},
  doi     = {10.1103/PhysRevA.94.062113}
}

@article{Rigas:2017,
  author  = {Rigas, I. and Kastrup, H. A. and M{\o}lmer, K. and S{\'a}nchez-Soto, L. L. and Leuchs, G.},
  title   = {Wigner functions for angle and orbital angular momentum},
  journal = {Phys. Rev. A},
  volume  = {95},
  pages   = {052111},
  year    = {2017},
  doi     = {10.1103/PhysRevA.95.052111}
}

@article{Hudson:1974,
  author  = {Hudson, R. L.},
  title   = {When is the {Wigner} quasi-probability density non-negative?},
  journal = {Rep. Math. Phys.},
  volume  = {6},
  pages   = {249--252},
  year    = {1974},
  doi     = {10.1016/0034-4877(74)90007-X}
}

@article{Soto:1983,
  author  = {Soto, F. and Claverie, P.},
  title   = {When is the {Wigner} function of multidimensional systems nonnegative?},
  journal = {J. Math. Phys.},
  volume  = {24},
  pages   = {97--100},
  year    = {1983},
  doi     = {10.1063/1.525607}
}

@article{Takagi:2018,
  author  = {Takagi, R. and Zhuang, Q.},
  title   = {Convex resource theory of non-{G}aussianity},
  journal = {Phys. Rev. A},
  volume  = {97},
  pages   = {062337},
  year    = {2018},
  doi     = {10.1103/PhysRevA.97.062337}
}

@article{Walschaers:2017,
  author  = {Walschaers, M. and Fabre, C. and Parigi, V. and Treps, N.},
  title   = {Entanglement and {W}igner Function Negativity of Multimode Non-{G}aussian States},
  journal = {Phys. Rev. Lett.},
  volume  = {119},
  pages   = {183601},
  year    = {2017},
  doi     = {10.1103/PhysRevLett.119.183601}
}

@article{Chabaud:2021,
  author  = {Chabaud, U.},
  title   = {Witnessing {W}igner Negativity},
  journal = {Quantum},
  volume  = {5},
  pages   = {471},
  year    = {2021},
  doi     = {10.22331/q-2021-06-08-471}
}

@article{fabre_wigner_2020,
	title = {Wigner distribution on a double cylinder phase space for studying quantum error correction protocol},
	volume = {102},
	rights = {All rights reserved},
	issn = {2469-9926, 2469-9934},
	url = {http://arxiv.org/abs/2005.09328},
	doi = {10.1103/PhysRevA.102.022411},
	pages = {022411},
	number = {2},
	journaltitle = {Physical Review A},
	journal = {Phys. Rev. A},
	author = {Fabre, N. and Keller, A. and Milman, P.},
	urldate = {2024-03-22},
	year= {2020},
	langid = {english},
	eprinttype = {arxiv},
	eprint = {2005.09328 [quant-ph]},
}

@article{mukunda_wigner_1979,
	title = {Wigner distribution for angle coordinates in quantum mechanics},
	volume = {47},
	issn = {0002-9505, 1943-2909},
	url = {http://aapt.scitation.org/doi/10.1119/1.11869},
	doi = {10.1119/1.11869},
	pages = {182--187},
	number = {2},
	journal= {American Journal of Physics},
	author = {Mukunda, N.},
	urldate = {2020-02-21},
	year = {1979},
	langid = {english},
}

@article{Rigas:2011,
  author  = {Rigas, I. and S{\'a}nchez-Soto, L. L. and Klimov, A. B. and {\v R}eh{\'a}{\v c}ek, J. and Hradil, Z.},
  title   = {Orbital angular momentum in phase space},
  journal = {Ann. Phys. (N.Y.)},
  volume  = {326},
  pages   = {426--439},
  year    = {2011},
  doi     = {10.1016/j.aop.2010.11.015}
}

@article{Aremua:2012,
  author  = {Aremua, I. and Gazeau, J.-P. and Hounkonnou, M. N.},
  title   = {Action-angle coherent states for quantum systems with cylindric phase space},
  journal = {J. Phys. A: Math. Theor.},
  volume  = {45},
  pages   = {335302},
  year    = {2012},
  doi     = {10.1088/1751-8113/45/33/335302}
}

@article{Liu:2017,
  author  = {Liu, H.-D. and Fu, L.-B. and Wang, X.-G.},
  title   = {Coherent-state approach for {M}ajorana representation},
  journal = {Commun. Theor. Phys.},
  volume  = {67},
  pages   = {611},
  year    = {2017},
  doi     = {10.1088/0253-6102/67/6/611}
}

@article{Chryssomalakos:2018,
  author  = {Chryssomalakos, C. and Guzm{\'a}n-Gonz{\'a}lez, E. and Serrano-Ens{\'a}stiga, E.},
  title   = {Geometry of spin coherent states},
  journal = {J. Phys. A: Math. Theor.},
  volume  = {51},
  pages   = {165202},
  year    = {2018},
  doi     = {10.1088/1751-8121/aab4cf}
}

@article{Fresneda:2018,
  author  = {Fresneda, R. and Gazeau, J.-P. and Noguera, D.},
  title   = {Quantum localisation on the circle},
  journal = {J. Math. Phys.},
  volume  = {59},
  pages   = {052105},
  year    = {2018},
  doi     = {10.1063/1.5021063}
}

@article{Kowalski:2021,
  author  = {Kowalski, K. and Lawniczak, K.},
  title   = {Wigner functions and coherent states for the quantum mechanics on a circle},
  journal = {J. Phys. A: Math. Theor.},
  volume  = {54},
  pages   = {275302},
  year    = {2021},
  doi     = {10.1088/1751-8121/ac03d0}
}

@article{Lutkenhaus:1995,
  author  = {L\"utkenhaus, N. and Barnett, S. M.},
  title   = {Non-classical effects in phase space},
  journal = {Phys. Rev. A},
  volume  = {51},
  pages   = {3340--3342},
  year    = {1995},
  doi     = {10.1103/PhysRevA.51.3340}
}

@article{Glauber:1963,
  author  = {Glauber, R. J.},
  title   = {Coherent and Incoherent States of the Radiation Field},
  journal = {Phys. Rev.},
  volume  = {131},
  pages   = {2766--2788},
  year    = {1963},
  doi     = {10.1103/PhysRev.131.2766}
}

@article{Goldberg:2020aa,
	author = {Goldberg,A. Z. and Klimov,A. B. and Grassl,M. and Leuchs,G. and S{\'a}nchez-Soto,L. L.},
	da = {2020/12/01},
	date = {2020/11/17},
	doi = {10.1116/5.0025819},
	journal = {AVS Quantum Sci.},
	m3 = {doi: 10.1116/5.0025819},
	number = {4},
	pages = {044701},
	title = {Extremal quantum states},
	ty = {JOUR},
	url = {https://doi.org/10.1116/5.0025819},
	volume = {2},
	year = {2020},
	year1 = {2020}}

@article{Nieto:1998aa,
	author = {Nieto, Luis Miguel and Atakishiyev, Natig M and Chumakov, Sergey M and Wolf, Kurt Bernardo},
	doi = {10.1088/0305-4470/31/16/015},
	issn = {0305-4470, 1361-6447},
	journal = {J. Phys. A: Math. Gen.},
	journaltitle = {Journal of Physics A: Mathematical and General},
	langid = {english},
	number = {16},
	pages = {3875--3895},
	title = {Wigner distribution function for {Euclidean} systems},
	url = {https://iopscience.iop.org/article/10.1088/0305-4470/31/16/015},
	urldate = {2022-02-02},
	volume = {31},
	year = {1998}}

@article{PhysRevLett.116.130402,
  title = {Wigner Distribution of Twisted Photons},
  author = {Mirhosseini, Mohammad and Maga\~na-Loaiza, Omar S. and Chen, Changchen and Hashemi Rafsanjani, Seyed Mohammad and Boyd, Robert W.},
  journal = {Phys. Rev. Lett.},
  volume = {116},
  issue = {13},
  pages = {130402},
  numpages = {6},
  year = {2016},
  month = {Apr},
  publisher = {American Physical Society},
  doi = {10.1103/PhysRevLett.116.130402},
  url = {https://link.aps.org/doi/10.1103/PhysRevLett.116.130402}
}

@article{Rigas:2011ut,
	author = {Rigas, I. and Sanchez-Soto, L. L. and Klimov, A. B. and Rehacek, J. and Hradil, Z.},
	doi = {10.1016/j.aop.2010.11.016},
	eprint = {1011.6184},
	eprinttype = {arxiv},
	issn = {00034916},
	journal = {Ann. Phys.},
	journaltitle = {Annals of Physics},
	langid = {english},
	number = {2},
	pages = {426--439},
	title = {Orbital angular momentum in phase space},
	urldate = {2020-01-16},
	volume = {326},
	year = {2011}}

@article{potocek_exponential_2015,
	title = {On the exponential form of the displacement operator for different systems},
	volume = {90},
	issn = {0031-8949, 1402-4896},
	url = {http://stacks.iop.org/1402-4896/90/i=6/a=065208?key=crossref.b0ceb2a75ae0a455aff328aab7ad5b18},
	doi = {10.1088/0031-8949/90/6/065208},
	pages = {065208},
	number = {6},
	journal = {Physica Scripta},
	author = {Potoček, Václav and Barnett, Stephen M},
	urldate = {2018-06-25},
	year = {2015},
	langid = {english},
}

@article{descamps_superselection_2024,
	title = {Superselection rules and bosonic quantum computational resources},
	volume = {133},
	rights = {All rights reserved},
	issn = {0031-9007, 1079-7114},
	url = {http://arxiv.org/abs/2407.03138},
	doi = {10.1103/PhysRevLett.133.260605},
	pages = {260605},
	number = {26},
	journaltitle = {Physical Review Letters},
	journal = {Phys. Rev. Lett.},
	author = {Descamps, Eloi and Fabre, Nicolas and Saharyan, Astghik and Keller, Arne and Milman, Pérola},
	urldate = {2025-02-15},
	year= {2024},
	langid = {english},
	eprinttype = {arxiv},
	eprint = {2407.03138 [quant-ph]},
}

@article{chabaud_resources_2023,
	title = {Resources for bosonic quantum computational advantage},
	volume = {130},
	issn = {0031-9007, 1079-7114},
	url = {http://arxiv.org/abs/2207.11781},
	doi = {10.1103/PhysRevLett.130.090602},
	pages = {090602},
	number = {9},
	journaltitle = {Physical Review Letters},
	journal = {Phys. Rev. Lett.},
	author = {Chabaud, Ulysse and Walschaers, Mattia},
	urldate = {2023-05-14},
	year = {2023},
	langid = {english},
	eprinttype = {arxiv},
	eprint = {2207.11781 [quant-ph]},
}

@article{walschaers_non-gaussian_2021,
	title = {Non-Gaussian Quantum States and Where to Find Them},
	volume = {2},
	issn = {2691-3399},
	url = {https://link.aps.org/doi/10.1103/PRXQuantum.2.030204},
	doi = {10.1103/PRXQuantum.2.030204},
	pages = {030204},
	number = {3},
	journaltitle = {{PRX} Quantum},
	journal = {{PRX} Quantum},
	author = {Walschaers, Mattia},
	urldate = {2025-02-12},
	year = {2021},
	langid = {english},
}

@article{Rigas:2010um,
	author = {Rigas, I. and S{\'a}nchez-Soto, L. L. and Klimov, A. B. and {\v R}eh{\'a}{\v c}ek, J. and Hradil, Z.},
	doi = {10.1103/PhysRevA.81.012101},
	issn = {1050-2947, 1094-1622},
	journal = {Phys. Rev. A},
	journaltitle = {Physical Review A},
	langid = {english},
	number = {1},
	pages = {012101},
	title = {{Non-negative Wigner functions for orbital angular momentum states}},
	url = {https://link.aps.org/doi/10.1103/PhysRevA.81.012101},
	urldate = {2020-01-16},
	volume = {81},
	year = {2010}}

@article{5fl9-89j4,
  title = {Non-Gaussianity from Superselection Rules},
  author = {Moulonguet, Nicolas and Descamps, Eloi and Lorger\'e, Jos\'e and Saharyan, Astghik and Keller, Arne and Milman, P\'erola},
  journal = {Phys. Rev. Lett.},
  volume = {137},
  issue = {5},
  pages = {050203},
  numpages = {8},
  year = {2026},
  month = {Jul},
  publisher = {American Physical Society},
  doi = {10.1103/5fl9-89j4},
  url = {https://link.aps.org/doi/10.1103/5fl9-89j4}
}

@article{Bjork:2015ux,
	author = {Bjork, G. and Klimov, A. B. and de la Hoz, P. and Grassl, M. and Leuchs, G. and S{\'a}nchez-Soto, L. L.},
	doi = {10.1103/PhysRevA.92.031801},
	eprinttype = {arxiv},
	issn = {1050-2947, 1094-1622},
	journal = {Phys. Rev. A},
	number = {3},
	pages = {031801},
	title = {{Extremal quantum states and their Majorana constellations}},
	urldate = {2022-02-09},
	volume = {92},
	year = {2015}}

@book{Gazeau:2009vb,
	address = {New York},
	author = {Gazeau, J.-P.},
	langid = {english},
	pages = {360},
	publisher = {Wiley},
	title = {{Coherent States in Quantum Physics}},
	year = {2009}}

@article{Chabaud:2020th,
	author = {Chabaud, U. and Markham, D. and Grosshans, F.},
	doi = {10.1103/PhysRevLett.124.063605},
	issn = {0031-9007, 1079-7114},
	journal = {Phys. Rev. Lett.},
	journaltitle = {Physical Review Letters},
	langid = {english},
	number = {6},
	pages = {063605},
	title = {Stellar Representation of Non-{Gaussian} Quantum States},
	url = {https://link.aps.org/doi/10.1103/PhysRevLett.124.063605},
	urldate = {2022-01-26},
	volume = {124},
	year = {2020}}

\end{document}